\documentclass[aps,prb,amsmath,twocolumn]{revtex4}
\usepackage{graphicx}
\usepackage{bm}

\begin{document}
\title{Electron transport through interacting quantum dots}

\author{Dmitri S. Golubev$^{1,3}$, and
Andrei D. Zaikin$^{2,3}$}
\affiliation{$^1$Institut f\"ur Theoretische Festk\"orperphysik,
Universit\"at Karlsruhe, 76128 Karlsruhe, Germany \\
$^{2}$Forschungszentrum Karlsruhe, Institut f\"ur Nanotechnologie,
76021, Karlsruhe, Germany\\
$^{3}$I.E. Tamm Department of Theoretical Physics, P.N.
Lebedev Physics Institute, 119991 Moscow, Russia}

\begin{abstract}
We present a detailed theoretical investigation of the effect of 
Coulomb interactions on electron transport
through quantum dots and double barrier structures connected
to a voltage source via an arbitrary linear impedance. Combining real
time path integral techniques with the scattering matrix approach
we derive the effective action and evaluate the
current-voltage characteristics of quantum dots at sufficiently 
large conductances. Our analysis reveals a reach variety of different 
regimes which we specify in details for the case of chaotic 
quantum dots. At sufficiently low energies the interaction correction
to the current depends logarithmically on temperature and voltage. We 
identify two different logarithmic regimes with the crossover
between them occurring at energies of order of the inverse dwell time
of electrons in the dot. We also analyze the frequency-dependent shot
noise in chaotic quantum dots and elucidate its direct relation to interaction
effects in mesoscopic electron transport.

\end{abstract}
\maketitle

\section{Introduction}

Low temperature electron transport in disordered conductors  
with electron-electron interactions remains one of the most 
intriguing topics in modern mesoscopic physics. 
Interplay between charge discreteness, quantum coherence,
scattering and interactions yields a rich variety of non-trivial
effects, many of which can adequately be described only within
a rather complicated theoretical framework.  

One of the most successful theoretical approaches in 
mesoscopic physics is the scattering
matrix formalism \cite{Bue,been}. In the absence of interactions this method
allows for a complete and physically transparent analysis of
electron dynamics in coherent conductors which includes not
only electrical conductance but also shot noise \cite{bb}
and eventually all higher cumulants of the current operator
\cite{LLL}. Formally it is possible to generalize the
Landauer formalism to systems with interactions \cite{meir}.
Within this approach, however, the electron Green functions in the 
interacting region still remain to be evaluated diagrammatically or 
by other means. 

An alternative way to tackle the problems with disorder and
interactions is to employ non-perturbative path-integral-based
techniques \cite{GZ,KA}. Recently it was demonstrated 
\cite{Naz,GZ00,GGZ02} that the advantages
of the path integral methods and the scattering
matrix approach can be conveniently combined within one formalism, 
thereby providing a powerful tool to theoretically describe
interaction effects in mesoscopic conductors.
Several important results have already been obtained
in this way. For instance, with the aid of the instanton technique it was 
shown \cite{Naz} that -- similarly to tunnel junctions
\cite{pa91,WG} -- arbitrary coherent conductors can exhibit the
 phenomenon of weak charge quantization. Unfortunately, in the limit
 of large dimensionless conductances $g\gg 1$ this effect is rather
 weak and it gains importance only
 at exponentially small temperatures and voltages. Furthermore, 
weak charge quantization vanishes completely even at $T=0$ provided
at least one of the conducting channels is fully transparent 
\cite{Mat,Naz}. The results \cite{Naz} were later confirmed
and extended in Refs. \onlinecite{Kam} and \onlinecite{GZ02}.

Another -- much more robust -- effect of electron-electron interactions
is the so-called interaction correction to the $I-V$ curve. This
interaction correction is negative and its 
magnitude scales linearly with the parameter \cite{GZ00} 
\begin{equation}
\beta =\frac{\sum_nT_n(1-T_n)}{\sum_nT_n},
\label{beta}
\end{equation} 
where $T_n$ is the transmission of the $n$-th conducting channel.
In contrast to the effect of weak charge quantization, even for
large conductances $g \gg 1$ the interaction correction remains
clearly observable up to high temperatures and it vanishes only
for $\beta \to 0$, i.e. provided {\it all} the conducting channels in 
the system are fully transparent. Moreover, as both temperature
and voltage get lowered, the interaction correction grows logarithmically
\cite{GZ00,Levy}. As a result, at sufficiently low energies 
the smallness $\sim 1/g$ gets compensated by a large logarithm and
the system enters a non-perturbative regime where terms of all orders
in the interaction need to be evaluated. Let us also recall that
for a particular case of highly conducting tunnel junctions 
this logarithmic enhancement of the interaction
correction is already well known for single- and double-junction
systems \cite{many1,many2} as well as for tunnel junction arrays 
\cite{many3,many4}.   

It is interesting to observe that both interaction correction to
the conductance and the shot noise spectrum \cite{bb} 
are proportional to the same parameter
$\beta$ (\ref{beta}). This observation illustrates a close relation
between quantum noise and interaction effects in coherent conductors
\cite{GZ00,Levy}. Proceeding further along these lines one can 
investigate the effect of electron-electron interactions on current
noise \cite{GGZ02}. The interaction correction 
to the Nyquist noise was again found to scale with the 
parameter (\ref{beta}) (in accordance with the fluctuation-dissipation 
theorem) while the same correction to the
shot noise turned out to be proportional to the third cumulant of the current
operator. It was conjectured \cite{GGZ02} that the same
rule should apply for higher cumulants as well, i.e. the interaction
correction to the $n$-th cumulant should be proportional to the
$(n+1)$-th cumulant for all values of $n$. A general proof of this
conjecture was very recently provided in Refs. \onlinecite{Naz2,BN}.

Throughout the analysis \cite{Naz,GZ00,GGZ02} it was assumed that in the
absence of interactions the conductor is described by an 
{\it energy independent} (though otherwise general) scattering matrix
\cite{FN0}. While in a number of important cases the above assumption indeed
applies, in various other physical situations it turns
out to be insufficient. Therefore it would
be highly desirable to develop a generalization of the path integral
technique \cite{GZ00,GGZ02} to conductors described by energy
dependent scattering amplitudes. This generalization is the
primary goal of our present paper.

In physical terms our analysis should now effectively account for
internal dynamics of coherent conductors or, in other words,
for a finite {\it dwell time} of electrons. This effect should
be combined with that of electron-electron interactions. Below we will
accomplish this program for an important and widely studied
class of conductors -- the so-called quantum dots \cite{been,ABG}.
    
The structure of the paper and our main results are as follows.

In Sec. 2 we describe our general real time path integral formalism and
derive the effective action of interacting quantum dots,
Eqs. (\ref{Sdot}-\ref{h0}) of our paper. The effect of
electron-electron interactions is treated in a standard manner by
the Hubbard-Stratonovich decoupling of the Coulomb term in the Hamiltonian
and reducing the problem to that of an electron interacting with the
fluctuating quantum electromagnetic field defined on the Keldysh
contour. In order to handle the fermionic part of the problem we
combine our path integral analysis with the scattering matrix
approach. For the model of a quantum dot adopted here the latter
approach allows to {\it exactly} integrate out all the electron paths
and express the effective action only in terms of the fluctuating
fields which are then treated within an effective
quasiclassical approximation suitable for highly conducting quantum
dots with $g \gg 1$. 

In Sec. 3 we derive a general
expression for the current noise in chaotic quantum dots as a function
of frequency, voltage and temperature. Although this expression itself
does not include interactions, it turns out to be very useful for
better understanding of the relation between shot noise and
interaction effects in mesoscopic conductors. 

Sec. 4 is devoted to a detailed description of the current-voltage
characteristics of interacting quantum dots. In the most simple
voltage-biased limit the corresponding general expressions are 
presented in Sec. 4A by Eqs. (\ref{IL11})-(\ref{deltaGint}). 
Further analysis of these expressions for chaotic quantum dots 
in the leading non-trivial order in $1/g$ is carried out in Sec. 4B 
and 4C. A more general case of an arbitrary external impedance 
is considered in Sec. 4D. It is demonstrated that, as one goes
away from the voltage-biased limit, the interaction
correction to the current gets substantially modified and new
regimes become possible.  

Further discussion of our results and their comparison
to several recent experiments can be found in Sec. 5. 
In Appendices we present various technical details of our 
derivation of the effective action (Appendices A and B),
details of our averaging procedure (Appendix C) and general
expressions for the current in interacting quantum dots (Appendix D).


\section{Effective action and current operator}

\subsection{General formalism}
Let us consider a system of interacting electrons described by the Hamiltonian
\begin{eqnarray}
{ H}=
 \int d{\bm r}\, \psi^\dagger_{\sigma}({\bm r})
\left[-\frac{\nabla^2}{2m}-\mu +U({\bm r})\right] \psi_\sigma ({\bm r})
\hspace{1cm}
\nonumber\\
+\frac{1}{2} \int d{\bm r}\int d{\bm
r'}\, \psi^\dagger_{\sigma}({\bm r}) \psi^\dagger_{\sigma'}({\bm r'})
\frac{e^2}{|{\bm r}-{\bm r'}|} \psi_{\sigma'}({\bm
r'}) \psi_{\sigma}({\bm r}).
\label{ham}
\end{eqnarray}
Here the operator $\psi^\dagger_\sigma({\bm r})$ $\big(\psi_\sigma({\bm r})\big)$
creates (annihilates) an electron with the coordinate ${\bm r}$ and
the spin projection $\sigma$. In Eq. (\ref{ham}) the summation 
over $\sigma$ is assumed.  
The time evolution of the density matrix of the whole
system ${\bm \rho}(t)$ is determined by the equation
\begin{equation}
{\bm \rho}(t)={\rm T}\exp\left[-i{ H}t\right]{\bm \rho}(0)
\tilde{\rm T}\exp\left[i{ H}t\right].
\end{equation}
Applying the Hubbard-Stratonovich transformation we rewrite
this equation in the form
\begin{eqnarray}
{\bm \rho}(t)=\frac{\int{\cal D}V_j\;{\rm T}{\rm e}^{-i\int_0^t dt'{ H}_1(t')}
{\bm \rho}(0)
\tilde{\rm T}{\rm e}^{i\int_0^t dt'{ H}_2(t')}{\rm e}^{iS_{\rm em}}}
{\int{\cal D}V_j\,{\rm e}^{iS_{\rm em}}},
\label{rho}
\end{eqnarray}
where
\begin{equation}
S_{\rm em}=\sum_{j=1,2}(-1)^{j+1}\int_0^t dt'\int d{\bm r}
\frac{\big(\nabla V_j(t',{\bm r})\big)^2}{8\pi}.
\label{Sem}
\end{equation}
is the electromagnetic contribution to the action,
the symbols ${\rm T}$ and $\tilde{\rm T}$ imply ordering
respectively in the forward and backward time directions
and
\begin{equation}
{ H}_{j}= \int d{\bm r}\, \psi^\dagger_{\sigma}({\bm r})
\left[-\frac{\nabla^2}{2m}-\mu +U({\bm r})-eV_{j}(t',{\bm r})\right] \psi_\sigma ({\bm r}).
\nonumber
\end{equation}

As usually, the expectation value of any physical
operator is obtained by taking the trace of this operator multiplied
by ${\bm\rho}(t)$. Here we are mainly interested in the
expectation value of the current operator. This operator, which we
will specify later, can be obtained by taking the functional
derivatives of the combination
${\rm T}(\tilde{\rm T})\exp\left[\pm i\int_0^t dt'{H}_{1,2}(t')\right]$
with respect to the fluctuating fields $V_{1,2}(t).$ For this reason there
is no need to deal with the whole expression (\ref{rho}), the kernel
of the current operator can be obtained from the effective action
$S,$ which is defined by the equation
\begin{eqnarray}
{\rm e}^{iS}={\rm Tr}\left\{
{\rm T}{\rm e}^{-i\int_0^t dt'{ H}_1(t')}
{\bm \rho}(0)
\tilde{\rm T}{\rm e}^{i\int_0^t dt'{H}_2(t')}
\right\}
{\rm e}^{iS_{\rm em}}.
\end{eqnarray}
The effective action $S$ can be expressed through the Keldysh Green
function matrix $\check G^{-1}_{V_1,V_2},$
\begin{equation}
iS=2{\rm Tr}\ln\check G^{-1}_{V_1,V_2} +iS_{\rm em},
\end{equation}
and the matrix $\check G^{-1}_{V_1,V_2}$ satisfies the equation
\begin{eqnarray}
\left[\left(i\frac{\partial}{\partial t}+\frac{\nabla^2}{2m}+\mu -U({\bm r}) +eV^+(t,{\bm r})\right)
\check 1 + 
\delta\check G^{-1}\right]
\nonumber\\
\times\, \check G_{V_1,V_2}(t t',{\bm r} {\bm r}') 
=\check\sigma_z\delta(t-t')\delta({\bm r}-\bm{r}'), 
\label{G}
\end{eqnarray}
Here we defined $\delta\check G^{-1}=(eV^-(t',{\bm r})/2)\check \sigma_z$
(where $\check \sigma_z$ is one of the Pauli matrices) and introduced 
symmetric and anti-symmetric combinations of the fluctuating fields:
\begin{equation}
V^+=(V_1+V_2)/2,\;\; V^-=V_1-V_2.
\end{equation}
As we have already discussed, in the interesting for us limit
of large dot conductances fluctuations of the fields $V_{1,2}$
remain relatively small. In this case one can expand
the exact effective action in $V^-$ keeping only the first and the
second orders. Then one finds
\begin{eqnarray}
i{ S}=iS_{\rm em} +2{\rm Tr}[\check G_{V^+}\delta\check G^{-1}]
-\, {\rm Tr}[(\check G_{V^+}\delta\check G^{-1})^2] ,
\label{rasl}
\end{eqnarray}
where
$\check G_{V^+}$ is the solution of Eq. (\ref{G}) with
$\delta\check G^{-1}=0.$  From (\ref{rasl}) we obtain
\begin{eqnarray}
i S&=& iS_{\rm em} 
+ e\int_0^t dt'\int d^3{\bm r} \big(  G_{V^+,11}(t'-0,t',{\bm r},{\bm r})
\nonumber\\ &&
+\, G_{V^+,22}(t'+0,t',{\bm r},{\bm r})\big) V^-(t',{\bm r})
\nonumber\\ &&
-\,e^2 \int_0^t dt' \int_0^t dt''\int d^3{\bm r}' d^3{\bm r}''\,
G_{V^+,12}(t',t'',{\bm r'},{\bm r''})
\nonumber\\ &&
\times\,  {V}^-(t'',{\bm r}'') G_{V^+,21}(t'',t',{\bm r}'',{\bm r}')) {V}^-(t',{\bm r}').
\label{S1}
\end{eqnarray}
Note that at this stage it is important {\it not}
to expand in $V^+$ keeping the exact nonlinear dependence of (\ref{rasl})
on this field.

The components of the Green function $\check G_{V^+}$ can be expressed through
the initial single particle electron density matrix $\hat\rho_0$ and
the single particle evolution operator $\hat U^{\varphi^+}.$ These are
the matrices in the channel space denoted by a hat here and below.
The operator $\hat U^{\varphi^+}$ is evaluated in the Appendix A. We have
\begin{eqnarray}
\hat G_{V^+,11}(t_1,t_2)&=&-i\theta(t_1-t_2)\hat U^{\varphi^+}(t_1,t_2)
\nonumber\\ &&
+\,i\hat U^{\varphi^+}(t_1,0)\hat\rho_0\hat U_{\varphi^+}(0,t_2),
\nonumber\\
\hat G_{V^+,22}(t_1,t_2)&=&-i\theta(t_2-t_1)\hat U^{\varphi^+}(t_1,t_2)
\nonumber\\ &&
+\,i\hat U^{\varphi^+}(t_1,0)\hat\rho_0\hat U_{\varphi^+}(0,t_2),
\nonumber\\
\hat G_{V^+,12}(t_1,t_2)&=&i\hat U^{\varphi^+}(t_1,0)\hat\rho_0\hat U^{\varphi^+}(0,t_2),
\nonumber\\
\hat G_{V^+,21}(t_1,t_2)&=&-i\hat U^{\varphi^+}(t_1,0)[\hat 1-\hat\rho_0]\hat U^{\varphi^+}(0,t_2).
\label{Gij}
\end{eqnarray}
Here for future purposes we have defined
\begin{equation}
\varphi^\pm(t')=\int_0^{t'}dt'' eV^\pm(t'').
\label{phi}
\end{equation}


\subsection{Effective action}

\begin{figure}
\centerline{\includegraphics[width=8cm]{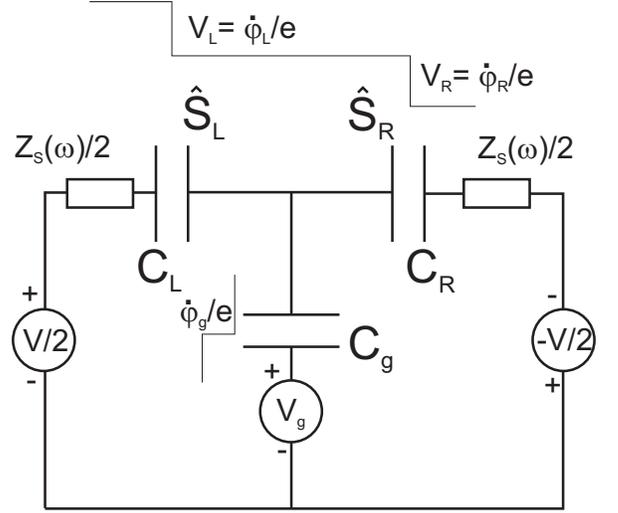}}
\caption{Schematics of the quantum dot.}
\end{figure}

The above analysis is rather general and can be applied to a variety
of mesoscopic structures with disorder and interactions. Our primary
goal here is to consider electron transport through a system which
contains an interacting quantum dot. This system is schematically displayed in Fig. 1.

A quantum dot can be viewed as an island in-between two barriers connected to a
voltage source via metallic leads with an arbitrary impedance $Z(\omega
)$. Electrons can enter the dot through one of the barriers, spend
some time there propagating between the barriers and
being reflected, and finally leave the dot through another barrier.
Details of the electron motion inside the dot will not be
important for us: Electrons can either propagate ballistically from
one barrier to the other or suffer additional scattering inside the
dot, e.g. from the outer walls or otherwise. 

Electron-electron interactions
are taken into account by means of the effective capacitance model. Here we
introduce capacitances of the left and right barriers $C_{L,R}$
and the gate capacitance $C_g$.
Making use of this model we can specify the expression for the
effective action further. Namely, we will assume that strong changes of the
fluctuating voltage fields $V^{\pm}$ in space are allowed only in the vicinity
of the barriers where they suffer jumps $V_L^{\pm}$ and $V_R^{\pm}$
(see Fig. 1). Additional voltage drop inside the dot is neglected,
i.e. there only the time dependence of the fields $V^{\pm}$
is taken into account. In the leads these fields
are assumed to vary slowly in space. 

Let us introduce the
fluctuating potentials of the left and right leads $V_{l,r}^\pm$
and of the dot $V_d^\pm$. Then we define the (time dependent) voltages 
across the barriers,
$V_{L}^{\pm}=V_l^\pm-V_d^\pm$ and $V_{R}^{\pm}=V_d^\pm-V_r^\pm$, and
find the electromagnetic potentials in the rest of our system by 
minimizing the action
(\ref{S1}). Afterwards we substitute the result back into
Eq. (\ref{S1}) and arrive at the expression which depends only on the
phase jumps $\varphi^\pm_{L,R}(t')=\int_0^{t'}dt'' eV^\pm_{L,R}(t'').$
In this way two contributions to the total effective action can be identified,
\begin{equation}
iS=iS_{\rm dot}+iS_{\rm ext}.
\end{equation}
Here the term $iS_{\rm dot}$ comes from the last two terms of Eq. (\ref{S1})
where the space integrals run over the inner part of the dot
and the barrier area. The second contribution $iS_{\rm ext}$
comes from the integrals over the remote parts of the system (leads).
Making use of a slow coordinate dependence of the fields $V^\pm$
in the leads one can expand the action
$iS_{\rm ext}$ in these fields and keep the first and the second order terms in
this expansion. We arrive at the standard form of the action \cite{SZ}
describing fluctuations produced by an arbitrary environment with a
linear impedance. Above we introduced the two leads with identical impedances $Z_S(\omega)/2.$
The fluctuating part of the voltage across the left lead is $(\dot\varphi^+_L-\dot\varphi^+_g)/e,$
the corresponding value for the right lead is $(\dot\varphi^+_R+\dot\varphi^+_g)/e.$
The corresponding contributions to the action need to be added 
and we arrive at the following expression 
\begin{widetext}
\begin{eqnarray}
iS_{\rm ext}&=&-\frac{i}{e^2}\int_0^t dt'\left(C_L\ddot\varphi^+_L\varphi^-_L+C_R\ddot\varphi^+_R\varphi^-_R
+C_g\ddot\varphi^+_g\varphi^-_g\right)
\nonumber\\
&&
-\,\frac{2i}{e^2}\int_0^tdt_1dt_2\, Z^{-1}_S(t_1-t_2)
\left[(\varphi^-_L(t_1)-\varphi^-_g(t_1))(\dot\varphi^+_L(t_2)-\dot\varphi^+_g(t_2))
+(\varphi^-_R(t_1)+\varphi^-_g(t_1))(\dot\varphi^+_R(t_2)+\dot\varphi^+_g(t_2))\right]
\nonumber\\
&&
-\,\frac{1}{e^2}\int_0^tdt_1dt_2\, {\cal Z}_S(t_1-t_2)
\left[(\varphi^-_L(t_1)-\varphi^-_g(t_1))(\varphi^-_L(t_2)-\varphi^-_g(t_2))
+(\varphi^-_R(t_1)+\varphi^-_g(t_1))(\varphi^-_R(t_2)+\varphi^-_g(t_2))\right],
\nonumber\\
&&
Z_S^{-1}(t)=\int\frac{d\omega}{2\pi}\,\frac{{\rm e}^{-i\omega t}}{Z_S(\omega)},\;\;\;
{\cal Z}_S(t)=\int\frac{d\omega}{2\pi}\,\omega\coth\frac{\omega}{2T}\,{\rm e}^{-i\omega t}\,
{\rm Re}\left(\frac{1}{Z_S(\omega)}\right).
\label{Sext}
\end{eqnarray}
\end{widetext}
The term $iS_{\rm em}$ is evaluated within the capacitance model and
included in the above expression for $iS_{\rm ext}$ (\ref{Sext}).

Now let us specify the effective action of the quantum dot. 
In what follows we will assume that the left (right) lead
contains $N_L$ ($N_R$) conducting channels. 
Every channel is characterized by the velocity
$v_n$ (which may also be different in different leads). In the absence
of electron-electron interactions quantum transport through the dot
can be fully described by the scattering matrix \cite{Bue,been}. As we have
already pointed out in Sec. 1 a convenient way
to include interaction effects is to combine the scattering matrix
approach with the path integral technique. 
As compared to Refs. \onlinecite{Naz,GZ00,GGZ02}
here we are dealing with a more complicated situation because
the scattering matrix of the dot can now depend on the energy $E$
of incoming electrons. Therefore, the action
\cite{GZ00,GGZ02} cannot be directly used and a proper generalization
is required for our problem.

Let us define the scattering matrix describing electron transport
through our quantum dot in the absence of interactions:
\begin{equation}
\hat S(E)=\left(\begin{array}{cc}
\hat r(E) & \hat t'(E) \\
\hat t(E) & \hat r'(E)
\end{array}\right).
\end{equation}
For our derivation it is also convenient to perform the Fourier 
transformation of the $S-$matrix. We introduce
$$\hat S(t)=\int\frac{dE}{2\pi}\hat S(E){\rm e}^{-iEt},\;\;\;\;
\hat S^\dagger(t)=\int\frac{dE}{2\pi}\hat S^\dagger(E){\rm e}^{iEt},$$
and the functions
$\hat r(t),$ $\hat r'(t),$ $\hat t(t),$ $\hat t'(t)$ defined analogously.
In accordance with the causality principle one has
$\hat S(t<0)=\hat S^\dagger(t<0)\equiv 0$.

It turns out that one can find an explicit expression for the
evolution operator $\hat U^{\varphi^+}$ in terms of the scattering
matrix, construct the Green functions (\ref{Gij}) and then derive
the effective action  of the quantum dot $S_{\rm dot}$ from
Eq. (\ref{S1}). This program
is carried out in Appendices A and B. The final result can be
expressed in the form
\begin{equation}
iS_{\rm dot}=iS_R-S_I.
\label{Sdot}
\end{equation}
As usually, the part $S_R$ describes dissipative effects. It reads
(see Appendix B):
\begin{eqnarray}
iS_R&=&-2i\int_0^t dz\int_0^\infty dx dy\,
{\rm tr}\big\{[\delta(z-y)\hat\varphi^-(z)\delta(z-x)
\nonumber\\ &&
-\,
\hat S^\dagger(z-y)\hat\varphi^-(z)\hat S(z-x)]\hat\rho(y,x) \big\}.
\label{SRcompact}
\end{eqnarray}
where
\begin{eqnarray}
\hat\varphi^-(z)=\left(
\begin{array}{c}
-\hat 1\,\varphi^-_L(z) \hspace{0.7cm} 0 \hspace{0.7cm} \\
\hspace{0.7cm} 0 \hspace{0.7cm} \hat 1\,\varphi^-_R(z)
\end{array}
\right),\hspace{2.4cm}
\nonumber\\
\hat\rho(y,x)=\rho_0(y-x)\left(
\begin{array}{c}
\hat 1\,{\rm e}^{i[\varphi^+_L(y)-\varphi^+_L(x)]} \hspace{0.7cm} 0 \hspace{0.7cm} \\
\hspace{0.7cm} 0 \hspace{0.7cm} \hat 1\,{\rm e}^{i[\varphi^+_R(x)-\varphi^+_R(y)]}
\end{array}
\right)
\label{rr0}
\end{eqnarray}
($\hat 1$ is the unity matrix), and
\begin{equation}
\rho_{0}(x)=\int\frac{dE}{2\pi}\,\frac{{\rm e}^{iEx}}{1+{\rm e}^{E/T}}
=\frac{1}{2}\delta(x)-\frac{iT}{2\sinh[\pi Tx]}
\label{rhoLR}
\end{equation}
is the equilibrium density matrix of non-interacting electrons.

The second term $S_I$ in (\ref{Sdot}) accounts for quantum noise. This
term is also evaluated in Appendix B. As a result we obtain
\begin{eqnarray}
S_I&=& \int_0^t dx_1dx_2\int_0^\infty dy_1dy_2\int_0^\infty dz_1 dz_2\; 
\nonumber\\ &&\times\,
{\rm tr}\big\{
\big[\delta(x_1-z_1)\hat\varphi^-(x_1)\delta(x_1-y_1)
\nonumber\\ &&
-\,\hat S^\dagger(x_1-z_1)\hat\varphi^-(x_1) \hat S(x_1-y_1)\big]
 \hat \rho(y_2,y_1)
\nonumber\\
&&
\times\,\big[\delta(x_2-y_2)\hat\varphi^-(x_2)\delta(x_2-z_2)
\nonumber\\ &&
-\hat S^\dagger(x_2-y_2)\hat\varphi^-(x_2) \hat S(x_2-z_2)\big]
\hat h(z_1,z_2)
\big\}.
\label{SIcompact}
\end{eqnarray}
Here we defined
\begin{eqnarray}
\hat h(z_1,z_2)=h_0(z_1-z_2)\left(
\begin{array}{c}
{\rm e}^{i[\varphi^+_L(z_1)-\varphi^+_L(z_2)]} \hspace{0.4cm} 0 \hspace{0.4cm} \\
\hspace{0.4cm} 0 \hspace{0.4cm} {\rm e}^{i[\varphi^+_R(z_2)-\varphi^+_R(z_1)]}
\end{array}
\right),
\label{h0}
\end{eqnarray}
and $h_0( z)=\delta( z)-\rho_0( z)$.
We also note that in the long time limit the integration over
$x,y$ in Eq. (\ref{SRcompact}) and over $x_{1,2},$ $y_{1,2}$
in Eq. (\ref{SIcompact}) can be extended to the interval
$(-\infty,+\infty).$

The expressions (\ref{Sdot}-\ref{h0}) for the effective action of
the quantum dot represent the main technical result of our
paper. This effective action is defined by essentially nonlocal in
time expressions which account for electron-electron interaction
effects in the presence of a non-zero dwell time $\tau_D$ of electrons
in the quantum dot. Should $\tau_D$ be 
much shorter than any other relevant time scale in our problem,
the time dependence of the scattering
matrices $\hat S(t)$ and $\hat S^+(t)$ can be approximated by the 
$\delta$-function,  $\hat S(t)\propto \hat S^+(t) \propto \delta (t)$,
in which case the action (\ref{Sdot}-\ref{h0})
reduces to one derived in Ref. \onlinecite{GZ00}.

\subsection{Averaging of the action}

If one is not interested in mesoscopic fluctuations of the effective
action one can simplify the above expressions by
averaging Eqs.  (\ref{SRcompact}) and (\ref{SIcompact}) over
energy intervals exceeding the dot level spacing $\delta$.
Consider first the term $S_R$ (\ref{SRcompact}).
Let us illustrate the main idea by treating the average
${\rm tr}\langle\hat t^\dagger(x)\hat t(y) \rangle.$ We have
\begin{eqnarray}
{\rm tr}\langle\hat t^\dagger(x)\hat t(y)\rangle=
\int\frac{dE_{1,2}}{(2\pi)^2} {\rm tr}\langle \hat t^\dagger(E_1)\hat t(E_2)\rangle
{\rm e}^{iE_1x-iE_2y}.
\end{eqnarray}
In a broad interval of energies the average
${\rm tr}\langle \hat t^\dagger(E_1)\hat t(E_2)\rangle$
should depend only the energy difference $E_1-E_2.$
Making use of this observation let us define the function $u^{RL}_\omega={\rm tr}\langle \hat t^\dagger(E)\hat t(E+\omega)\rangle$
and its Fourier transform $u_{RL}(t)=\int\frac{d\omega}{2\pi}\,u^{RL}_\omega\,{\rm e}^{-i\omega t}.$
The function $u^{RL}_\omega$ satisfies the property
$u^{RL}_{-\omega}={u^{RL}_\omega}^*,$ therefore
$u_{RL}(t)$ is real. Other averages are defined analogously. We find
\begin{eqnarray}
{\rm tr}\langle[\delta(x)\delta(y)\hat 1- \hat r^\dagger(x)\hat r(y)]\rangle&=&
\delta(x-y)u_{LL}(y),
\nonumber\\
{\rm tr}\langle\hat {t'}^\dagger(x)\hat t'(y) \rangle&=&
\delta(x-y)u_{LR}(y),
\nonumber\\
{\rm tr}\langle[\delta(x)\delta(y)\hat 1- \hat {r'}^\dagger(x)\hat r'(y)]\rangle&=&
\delta(x-y)u_{RR}(y),
\nonumber\\
{\rm tr}\langle\hat t^\dagger(x)\hat t(y)\rangle &=&
\delta(x-y)u_{RL}(y).
\label{uuu}
\end{eqnarray}
Averaging of Eq. (\ref{SRcompact}) with the aid of (\ref{uuu})
yields
\begin{eqnarray}
iS_R^{\rm av}=
-\frac{i}{\pi}\sum_{i,j=L,R}\int\limits_0^t dz\int\limits_0^z dx\,
\varphi^-_i(z)u_{ij}(z-x)\dot\varphi_j^+(x).
\label{SRav}
\end{eqnarray}   
We note that $S_R^{\rm av}$ -- unlike the non-averaged action
$S_R$ (\ref{SRcompact}) -- is bilinear in both $\varphi^+$ and
$\varphi^-.$

Now let us average the term $S_I$. We first notice that, since this
term is already quadratic $\varphi^-$, in the averaged version of
$S_I$ one can neglect fluctuations of the phases
$\varphi^\pm_{L,R}$ and set $\varphi^+_{L,R}(t)=eV_{L,R}t.$
After that the voltages $V_{L,R}$ can be absorbed as energy shifts
of the Fermi distribution functions in the leads.
Averaging of the components of the scattering matrix entering into
Eq. (\ref{SIcompact}) is carried out as above. We first
proceed to the energy representation. Then we will get a sum of terms
containing combinations of a similar structure, such as, e.g.,
\begin{eqnarray}
\int \frac{dE}{2\pi}f(E+\omega -eV_L)(1-f(E-eV_L))\hspace{1.8cm}
\nonumber\\
\times{\rm tr}\left\langle[\hat 1-\hat r^\dagger(E)\hat r(E+\omega)][\hat
  1-\hat r^\dagger(E+\omega)\hat r(E)]\right\rangle ,
\end{eqnarray}
where $f(E)=(1+\exp (E/T))^{-1}$ is the Fermi function. Since the averages 
should not depend on $E,$ in all these combinations
one can integrate over this variable. Collecting all terms we arrive
at the final result
\begin{eqnarray}
S_I^{\rm av}=\sum_{i,j}\int_0^t dx_1\int_0^t dx_2\,
\varphi^-_i(x_1)v_{ij}(x_1-x_2)\varphi_j^-(x_2).
\label{SIav}
\end{eqnarray}
Here the kernels $v$ are defined as
$v_{ij}(t)=\int\frac{d\omega}{2\pi}\,v^{ij}_\omega\,{\rm e}^{-i\omega
  t}$, where
\begin{eqnarray}
v^{ij}_\omega&=&\frac{\omega}{2\pi}\coth\frac{\omega}{2T}\,{\rm Re}\,u^{ij}_\omega+
\left[\frac{\omega-eV}{4\pi}\coth\frac{\omega-eV}{2T}
\right.
\nonumber\\ &&
\left.
+\,\frac{\omega+eV}{4\pi}\coth\frac{\omega+eV}{2T}
-\frac{\omega}{2\pi}\coth\frac{\omega}{2T}\right]
\tilde v^{ij}_\omega,
\label{vij}
\end{eqnarray}
$V=V_L+V_R$ and
\begin{eqnarray}
\tilde v^{LL}_\omega&=&{\rm tr}\langle \hat r(E)\hat r^\dagger(E)\hat t'(E+\omega)\hat{t'}^\dagger(E+\omega)\rangle,
\nonumber\\
\tilde v^{RL}_\omega&=&-{\rm tr}\langle \hat r(E)\hat t^\dagger(E)\hat r'(E+\omega)\hat{t'}^\dagger(E+\omega)\rangle,
\nonumber\\
\tilde v^{LR}_\omega&=&-{\rm tr}\langle \hat t(E)\hat r^\dagger(E)\hat t'(E+\omega)\hat{r'}^\dagger(E+\omega)\rangle,
\nonumber\\
\tilde v^{RR}_\omega&=&{\rm tr}\langle \hat t(E)\hat t^\dagger(E)\hat r'(E+\omega)\hat{r'}^\dagger(E+\omega)\rangle.
\label{dvij}
\end{eqnarray}
 This concludes our
derivation of the effective action for interacting quantum dots.

\subsection{Current operator}

In order to complete our general analysis let us define the kernel of 
the current operator for our problem.
One can choose calculating the current either in
the left or in the right junction, obviously in the stationary limit 
the result should remain the same in both cases. One can also use a
symmetrized version of the current operator. One finds
\begin{equation}
I=\frac{\int{\cal D}\varphi^-{\cal D}\varphi^+ \,I(t,\varphi^\pm)\,{\rm e}^{iS_{\rm ext}+iS_R-S_I}}
{\int{\cal D}\varphi^-{\cal D}\varphi^+ \,{\rm e}^{iS_{\rm ext}+iS_R-S_I}},
\label{I1}
\end{equation}
where
\begin{eqnarray}
I(t,\varphi^\pm)&=&\frac{1}{2}\left(-ie\frac{\delta(iS_R-S_I)}
{\delta\varphi^-_L(t)}-ie\frac{\delta(iS_R-S_I)}
{\delta\varphi^-_R(t)}\right)
\nonumber\\
&=&-\frac{ie}{2}\left.\frac{\delta(iS_R-S_I)}
{\delta\left(\frac{\varphi^-_L(t)+\varphi^-_R(t)}{2}\right)}\right|_{\varphi_L^--\varphi_R^-={\rm const}}.
\label{I2}
\end{eqnarray}
Since here we are interested in a stationary situation,
the result for the current should not depend on time.
Therefore one can choose an arbitrary value of $t$ at which the above
functional derivative is taken, with the only requirement that $t$
should be sufficiently large. Inserting Eqs. (\ref{SRcompact}) and
(\ref{SIcompact}) into (\ref{I2}), 
and defining the matrix
$$
\hat\Lambda=\left(\begin{array}{cc}
\hat 1 & 0 \\
0 & -\hat 1
\end{array}\right),
$$
we obtain
\begin{equation}
I(t,\varphi^\pm)=I_0(t,\varphi^+)+\delta I(t,\varphi^\pm),
\label{opI}
\end{equation}
where
\begin{eqnarray}
I_0(t,\varphi^+)&=&e\int dxdy\,{\rm tr}\left\{[\delta(t-x)\hat\Lambda\delta(t-y)
\right.
\nonumber\\ &&
\left.
-\,\hat S^\dagger(t-x)\hat\Lambda\hat S(t-y)]\hat\rho(x,y)\right\},
\label{opI0}
\end{eqnarray}
and
\begin{widetext}
\begin{eqnarray}
\delta I(t,\varphi^\pm)&=&
 -\frac{ie}{2}\int_0^t dx\int_0^\infty dy_1dy_2\int_0^\infty dz_1 dz_2\; {\rm Tr}\left\{
\big[\delta(t-z_1)\hat\Lambda\delta(t-y_1)
-\hat S^\dagger(t-z_1)\hat\Lambda \hat S(t-y_1)\big]
 \hat \rho(y_2,y_1)
\right.
\nonumber\\
&&\left.
\times\,\big[\delta(x-y_2)\hat\varphi^-(x)\delta(x-z_2) 
-\hat S^\dagger(x-y_2)\hat\varphi^-(x) \hat S(x-z_2)\big]
\hat h(z_1,z_2)
\right\}
\nonumber\\
&&-\,\frac{ie}{2}\int_0^t dx\int_0^\infty dy_1dy_2\int_0^\infty dz_1 dz_2\; {\rm Tr}\left\{
\big[\delta(x-z_1)\hat\varphi^-(x)\delta(x-y_1)
-\hat S^\dagger(x-z_1)\hat\varphi^-(x) \hat S(x-y_1)\big]
 \hat \rho(y_2,y_1)
\right.
\nonumber\\
&&\left.
\times\,\big[\delta(t-y_2)\hat\Lambda\delta(t-z_2)
-\hat S^\dagger(t-y_2)\hat\Lambda \hat S(t-z_2)\big]
\hat h(z_1,z_2)
\right\}.
\label{opdI}
\end{eqnarray}
\end{widetext}

\section{Noise of a quantum dot}

The general expression for the term $S_I$ (\ref{SIcompact}) in the action
enables one to easily evaluate the current correlators 
at the barriers in the absence of interaction:
\begin{eqnarray}
{\cal S}_{ij}(\omega)=\int dt\,{\rm e}^{i\omega t}\langle\delta I_i(t)\delta I_j(0)+\delta I_j(0)\delta I_i(t)\rangle,
\end{eqnarray}
where $i,j=L,R.$ One finds
\begin{equation}
{\cal S}_{ij}(\omega)=2e^2\int dt\,{\rm e}^{i\omega t}\,
\frac{\delta^2 S_I[\varphi^-_\alpha,\varphi^+_{\beta}=eV_{\beta}t]}{\delta\varphi^-_i(t)\varphi^-_j(0)}.
\end{equation}
In this way one recovers the well 
known general expression for the noise
in terms of the scattering amplitudes \cite{bb}.
Averaging the result over mesoscopic fluctuations we get
\begin{eqnarray}
{\cal S}_{ij}(\omega)=
\frac{2e^2}{\pi}\big({\rm Re}\,u^{ij}_\omega-\tilde v^{ij}_\omega\big)\omega\coth\frac{\omega}{2T}
+\frac{e^2}{\pi}\tilde v^{ij}_\omega
\hspace{1.2cm}
\nonumber\\ 
\times
\left((\omega-eV)\coth\frac{\omega-eV}{2T}+(\omega+eV)\coth\frac{\omega+eV}{2T}\right)
\label{Sij}
\end{eqnarray}

Following the standard procedure\cite{BB} let us express the
scattering matrix of the dot in the form
\begin{equation}
\hat S(E)=\hat R + \hat T'[1-\hat U(E)\hat R']^{-1}\hat U(E)\hat T,
\label{ST}
\end{equation}
where the energy independent matrices
\begin{eqnarray}
\hat R=\left(\begin{array}{cc}
\hat r_L & 0 \\
0 & \hat r'_R
\end{array}\right), \;\;
\hat R'=\left(\begin{array}{cc}
\hat r'_L & 0 \\
0 & \hat r_R
\end{array}\right)
\label{rrr} 
\end{eqnarray}
and
\begin{eqnarray}
\hat T=\left(\begin{array}{cc}
\hat t_L & 0 \\
0 & \hat t'_R
\end{array}\right), \;\;
\hat T'=\left(\begin{array}{cc}
\hat t'_L & 0 \\
0 & \hat t_R
\end{array}\right)
\label{ttt}
\end{eqnarray}
account for scattering properties of the left and right barriers while
$\hat U(E)$ is the unitary matrix  which effectively describes 
scattering in the ``internal'' part of the dot.
Further analysis requires specifying the model for our
quantum dot. Here we will mainly address {\it chaotic}
quantum dots in which case $\hat U(E)$ belongs to the circular
ensemble. The averages
$u^{ij}_\omega$ and $v^{ij}_\omega$ are evaluated with the aid of 
the diagrammatic technique \cite{BB,Brouwer}. We obtain
\begin{eqnarray}
u^{LL,RR}_\omega&=&\frac{g_{L,R}}{2}-\frac{g_{L,R}^2}{2g(1-i\omega\tau_D)},
\nonumber\\ 
u^{LR}_\omega&=&u^{RL}_\omega=\frac{g_Lg_R}{2g(1-i\omega\tau_D)}.
\label{uav}
\end{eqnarray} 
The details of the derivation are presented in Appendix C.
Here we defined 
the dimensionless conductances of the left and
right barriers
$g_{L,R}=2{\rm tr}\langle\hat t_{L,R}^\dagger(E) t_{L,R}(E) \rangle=
2\pi/e^2R_{L,R}$, their sum $g=g_L+g_R,$ and the electron 
dwell time in the quantum dot $\tau_D=4\pi/g\delta.$ The general
expressions for $\tilde v^{ij}_\omega$
can also be obtained with the aid of the results derived in Appendix C. 
One finds
\begin{eqnarray}
\tilde v^{LL}_\omega &=&
\frac{g_L^2g_R}{2g^3}\left(g_L\frac{\omega^2\tau_D^2}{1+\omega^2\tau_D^2}+g_R\right)
\nonumber\\ &&
+\, \frac{g_Lg_R^2}{2g^4}\frac{(g_L+g_R)^2\omega^2\tau_D^2+g_R^2}{1+\omega^2\tau_D^2}\beta_L
\nonumber\\ &&
+\,\frac{g_L^4g_R}{2g^4(1+\omega^2\tau_D^2)}\beta_R,
\nonumber\\ 
\tilde v^{RL}_\omega  
&=&\frac{g_L^2g_R^2}{2g^3(1+\omega^2\tau_D^2)}
\nonumber\\ && 
+\,\frac{g_Lg_R^3(i\omega\tau_D g_L+(1+i\omega\tau_D)g_R)}{2g^4(1+\omega^2\tau_D^2)}\beta_L
\nonumber\\ &&
+\,\frac{g_L^3g_R((1-i\omega\tau_D )g_L-i\omega\tau_D g_R)}{2g^4(1+\omega^2\tau_D^2)}\beta_R.
\hspace{1cm}
\label{vijav}
\end{eqnarray}
Here we introduced the Fano factors of the barriers 
$$
\beta_{j}=\frac{{\rm tr}\big(\hat t_j\hat t_j^\dagger\hat r'_j\hat {r'}^\dagger_j\big)}{
{\rm tr}\big(\hat t_j \hat t_j^\dagger\big)},\;\; j=L,R.
$$ 
The functions $\tilde v^{RR}_\omega$ and 
$\tilde v^{LR}_\omega$ are recovered by interchanging the indices 
$L\leftrightarrow R$
in the expressions (\ref{vijav}) for $\tilde v^{LL}_\omega$
and $\tilde v^{RL}_\omega$.

In the low frequency limit $\omega\to 0$ one
finds $\tilde v^{LL}=\tilde v^{RR}=\tilde v^{LR}=\tilde v^{RL}$ and, hence,
${\cal S}_{LL}(0)={\cal S}_{RR}(0)={\cal S},$ where ${\cal S}$ is the noise
spectrum of the dot,
\begin{eqnarray}
{\cal S}=\frac{4T}{R_q}\left(\frac{g_Lg_R(g^2-g_Lg_R)}{g^3}-\frac{g_Lg_R^4\beta_L+g_L^4g_R\beta_R}{g^4}\right)
\nonumber\\ 
+\,\frac{2eV}{R_q}\coth\frac{eV}{2T}\left(\frac{g_L^2g_R^2}{g^3}+\frac{g_Lg_R^4\beta_L+g_L^4g_R\beta_R}{g^4}\right).
\hspace{0.3cm}
\label{shum}
\end{eqnarray}
Here and below $R_q=h/e^2$ is the quantum resistance unit.

In the leading approximation the dot conductance 
is $G_{0}= g_Lg_R/gR_q$. Hence, for chaotic quantum dots 
the total Fano factor
$\tilde\beta$ (defined in a standard manner as a ratio between 
the shot noise spectrum and its Schottky value $2eI$) reads
\begin{equation}
\tilde\beta=\frac{g_Lg_R}{(g_L+g_R)^2}+\frac{g_R^3\beta_L+g_L^3\beta_R}{(g_L+g_R)^3}.
\label{Fano}
\end{equation} 
We note that Eq. (\ref{shum}) agrees with the results for zero
frequency noise of chaotic cavities previously derived in different
limits by various authors \cite{bb,NSP}. It is also satisfactory to 
observe that in the
particular case of two diffusive conductors $\beta_L=\beta_R=1/3$ 
Eq. (\ref{Fano}) again yields $\tilde\beta=1/3$ for any $g_L$ and $g_R$.

Our analysis also allows to generalize the results for the current
noise in chaotic cavities to the case of finite frequencies. 
In order to find the noise correlator ${\cal S}(\omega)$ 
it is in general necessary to account for the barrier capacitances.
Assuming for simplicity $C_g=0$, we obtain
\begin{eqnarray}
{\cal S}(\omega)&=&|Y_L(\omega)|^2{\cal
  S}_{LL}(\omega)+|Y_R(\omega)|^2{\cal S}_{RR}(\omega)
\nonumber\\ &&
+\,2{\rm Re}\big(Y_L(\omega){\cal S}_{LR}(\omega)Y_R(\omega)\big),
\label{Somega}
\end{eqnarray}
where
\begin{eqnarray}
Y_L(\omega)=\frac{-i\omega C_R+\frac{e^2}{\pi}\big(u^{RR}_\omega-u^{RL}_\omega\big)}
{-i\omega (C_L+C_R)+\frac{e^2}{\pi}\big(u^{LL}_\omega+u^{RR}_\omega-u^{LR}_\omega-u^{RL}_\omega\big)},
\nonumber\\
Y_R(\omega)=\frac{-i\omega C_L+\frac{e^2}{\pi}\big(u^{LL}_\omega-u^{LR}_\omega\big)}
{-i\omega (C_L+C_R)+\frac{e^2}{\pi}\big(u^{LL}_\omega+u^{RR}_\omega-u^{LR}_\omega-u^{RL}_\omega\big)}
\nonumber
\end{eqnarray}
and $u_\omega^{ij}$ are defined in (\ref{uav}). The above equations
demonstrate that in a general case the noise spectrum (\ref{Somega})
depends on frequency in a complicated manner.  
However, for fully symmetric quantum dots, i.e. for $C_L=C_R,$
$g_L=g_R$ and $\beta_L=\beta_R=\beta,$ one find $Y_L(\omega)=Y_R(\omega)=1/2$.
In this case all frequency dependent contributions contained in 
the functions $\tilde v^{ij}_\omega$ (\ref{vijav}) 
cancel out and Eq. (\ref{Somega}) reduces to the standard form \cite{bb}
\begin{eqnarray}
{\cal S}(\omega)&=&2(1-\tilde\beta)G_0\omega\coth\frac{\omega}{2T}
\nonumber\\ &&
+\,\tilde\beta G_0 \sum_{\pm}(\omega\pm eV)\coth\frac{\omega\pm eV}{2T}
\label{Khlus}
\end{eqnarray}
with the Fano factor $\tilde\beta=(1+\beta)/4$ in accordance with  Eq. (\ref{Fano}).

\section{Current-voltage characteristics and conductance}

Let us now turn to the interaction effects.
In order to evaluate the current-voltage characteristics we should 
average the above expressions for the phase-dependent current
$I(t,\varphi^\pm)$ (\ref{opI})-(\ref{opdI}) over the fluctuating phase fields
$\varphi^\pm$. In addition to that -- provided one is
not interested in the effect of mesoscopic fluctuations -- one can also
average the result over such fluctuations. We proceed exactly in this
order and first perform averaging over the phase fluctuations.

\subsection{Averaging over fluctuating phase fields}

Let us combine Eqs. (\ref{opI})-(\ref{opdI})
with (\ref{I1}) and carry out functional integration over
$\varphi^\pm$. We notice that in the interesting for us limit of large
dot conductances it is parametrically justified to perform this
integration with the averaged effective action
$$
S^{\rm av}=S_{\rm ext}+S_R^{\rm av}+iS_I^{\rm av}
$$
instead of the exact one. This observation simplifies
the whole procedure enormously because
the action $S^{\rm av}$ is quadratic in $\varphi^\pm$. Hence, the
integrals become Gaussian and can be handled exactly.

Since in all the integrals only linear combinations of the phases
$\varphi^+$ enter into the exponent, it is convenient to first
integrate out this variable. This integration
yields functional $\delta$-functions. For example, one  
of the terms in Eq. (\ref{opdI}) contains 
the exponent ${\rm e}^{i[\varphi^+_L(y_2)-\varphi^+_L(y_1)+\varphi^+_R(z_2)-\varphi^+_R(z_1)]},$ in which case 
the averaging over $\varphi^+_{L,R}$ fixes the variables
$\varphi^-$ in the form
\begin{eqnarray}
\varphi^-_j(\tau)&=&-K_{jL}(z_1-\tau)+K_{jL}(z_2-\tau)-K_{jR}(y_1-\tau)
\nonumber\\ &&
+\,K_{jR}(y_2-\tau), \;\; j=L,R,g.
\label{varphi}
\end{eqnarray}
The Fourier transform of the functions $K_{ij}(t)$ reads
\begin{equation}
K^{ij}_\omega =\frac{e^2}{-i\omega +0}\, {\cal
A}_{ij}^{-1}(\omega), \label{Kij}
\end{equation}
where the matrix ${\cal A}(\omega)$ has the form
\begin{widetext}
\begin{equation}
{\cal A}(\omega)=\left(
\begin{array}{ccc}
-i\omega C_L + \frac{2}{Z_S(\omega)}+\frac{e^2}{\pi}u_{LL}(\omega )&
 \frac{e^2}{\pi}u_{RL}(\omega ) & -  \frac{2}{Z_S(\omega)} \\
\frac{e^2}{\pi}u_{LR}(\omega ) & -i\omega C_R +
\frac{2}{Z_S(\omega)}+\frac{e^2}{\pi}u_{RR}(\omega ) & \frac{2}{Z_S(\omega)} \\
- \frac{2}{Z_S(\omega)} & \frac{2}{Z_S(\omega)} & -i\omega C_g + \frac{4}{Z_S(\omega)}\\
\end{array}
\right).
\label{calA}
\end{equation}
\end{widetext}
Other terms of Eq. (\ref{opdI}) are treated analogously.
The above expressions hold for an arbitrary external impedance
$Z_S(\omega)$. Further general expressions for the current are presented
in Appendix D. 

Let us now focus our attention on the important limit
of vanishing external impedance $Z_S(\omega)\to 0$.  In this limit one finds
\begin{equation}
K^{ij}_\omega =K_\omega \left(
\begin{array}{ccc}
1 & -1 & 1 \\
-1 & 1 & -1 \\
1 & -1 & 1 \\
\end{array}
\right),
\end{equation}
where we defined
\begin{eqnarray}
K_\omega &=&\frac{e^2}{-i\omega +0}\, \frac{1}{-i\omega C_\Sigma
+\frac{e^2}{\pi}u(\omega)},
\label{K}\\
u(\omega)&=&u^{LL}_\omega +u^{RR}_\omega -u^{LR}_\omega
-u^{RL}_\omega 
\nonumber\\
&=& {\rm tr}\langle\hat 1 -\hat S^\dagger(E)\hat S(E+\omega)\rangle.
\label{u}
\end{eqnarray}
and $C_\Sigma=C_L+C_R+C_g.$ What remains is to integrate out the 
variable $\varphi^-$. Because of
the obtained functional $\delta$-functions this integration becomes
trivial as well. One should just substitute the trajectories
(\ref{varphi}) into $S_I^{\rm av}$, into the imaginary part of
$S_{\rm ext}$ and into the term $\delta I$. Taking 
the unitarity of the $S-$matrix into account, in the voltage-biased
limit $Z_S(\omega)\to 0$ one finds
\begin{equation}
I(V)=I_0(V)+\delta I(V).
\label{IL11}
\end{equation}
For the sake of convenience we present the expression for the term
$I_0(V)$ in two equivalent forms: 
\begin{eqnarray}
I_0(V)&=&2e\int dx\rho_0(x){\cal T}(x)\,{\rm e}^{-F(x)}\left({\rm e}^{ieV_Lx}-{\rm e}^{-ieV_Rx}\right)
\nonumber\\
&=&e\int\frac{dEd\omega}{(2\pi)^2} {\rm tr}\{[\hat\Lambda-\hat S^\dagger(E+\omega)\hat\Lambda\hat S(E+\omega)]
\hat\rho_E\}
\nonumber\\ &&
\times\,
\left(\int dx\,{\rm e}^{i\omega x-F(x)}\right).
\label{I01}
\end{eqnarray}
Here we have defined ${\cal T}(x)=\int\frac{dE}{2\pi}\,{\rm tr}[\hat
  t^\dagger(E)\hat t(E)]\,{\rm e}^{-iEx},$ and
\begin{equation}
\hat\rho_E=\left(
\begin{array}{cc}
\hat 1 f(E-eV_L) & 0\\
0 & \hat 1 f(E+eV_R) \\
\end{array}
\right),
\end{equation} 
where $f(E)=1/(1+\exp(E/T))$ is the Fermi function.

The interaction correction to the current $\delta I(V)$ takes the form
\begin{widetext}
\begin{eqnarray}
\delta I&=&
 -e{\rm Im}\int dx\int dy_1dy_2\int dz_1 dz_2\,{\cal F}(y_1,y_2,z_1,z_2)\,
\big(K(-y_1)-K(-z_1)\big)
\nonumber\\ &&\times\,{\rm tr}\big\{
\big[\delta(z_1+x)\hat\Lambda\delta(y_1+x)
-\hat S^\dagger(z_1+x)\hat\Lambda \hat S(y_1+x)\big]
\hat \rho_0(y_1-y_2)
\big[\delta(y_2)\delta(z_2) 
-\hat S^\dagger(y_2) \hat S(z_2)\big]
\hat h_0(z_2-z_1)
\big\},
\label{deltaI1}
\end{eqnarray}
\end{widetext}
where we have set
\begin{eqnarray}
{\cal F}(y_1,y_2,z_1,z_2)=
{\rm e}^{-F(y_1-y_2)-F(z_1-z_2)-F(y_1-z_1)}
\hspace{0.5cm}
\nonumber\\
\times\,
{\rm e}^{-F(y_2-z_2)+F(y_1-z_2)+F(y_2-z_1)}.
\label{calF}
\end{eqnarray}
In the limit $Z_S(\omega)\to 0$ we have ${\rm Re}(iS_{\rm ext})=0$
and the function $F(x)$ is simply equal to
$F(x)=S_I^{\rm av}$ evaluated for
$\varphi^-_L(\tau)=-\varphi^-_R(\tau)=K(x-\tau)-K(-\tau).$ In this
case we find
\begin{eqnarray}
F(x)&=&2\int\frac{d\omega}{2\pi}\, |K_\omega |^2(1-\cos\omega
x)(v^{LL}_\omega +v^{RR}_\omega
\nonumber\\ &&
 -\,v^{LR}_\omega -v^{RL}_\omega ),
\label{Ffull}
\end{eqnarray}
where the functions $v^{ij}_\omega$ are defined in Eqs. (\ref{vij})-(\ref{dvij}).

As we shall see below, for a wide range of parameters the function 
(\ref{calF}) in Eq. (\ref{deltaI1}) can be approximated by unity 
${\cal F} \to 1$. Under this approximation one can conveniently perform the
Fourier transformation and reduce the expression
(\ref{deltaI1}) to the form
\begin{eqnarray}
\delta I=e{\rm Im}\int\frac{dEd\omega}{(2\pi)^2}K_\omega
{\rm tr}\big\{[1-\hat S^\dagger(E)\hat S(E+\omega)]\hat h_{E+\omega}
\nonumber\\ \times\,
[\hat S^\dagger(E+\omega)\hat\Lambda\hat S(E+\omega)-\hat S^\dagger(E)\hat\Lambda\hat S(E)]
\hat\rho_E\big\},\;\;
\label{deltaI}
\end{eqnarray}
where $\hat h_E=\hat 1-\hat \rho_E.$ 

Employing the above general expressions one can also define the zero bias 
conductance of the dot. In doing so, we express the voltages $V_{L,R}$
in the form $V_{L,R}=(V\pm V_G)/2,$ where $V_G$ is the effective potential of the
dot tuned through the gate, and find the linear in $V$ correction to the current. 
The voltage $V_G$ can be removed by shifting the energy and 
performing a proper re-definition of the $S-$matrix. 
Thus we find $G=G_F+\delta G$, where   
\begin{eqnarray}
G_F&=&-\frac{e^2}{2}\int\frac{dEd\omega}{(2\pi)^2}\, {\rm Tr}\left\{\hat 1 -\hat S^+(E+\omega)\hat\Lambda\hat S(E+\omega)\hat\Lambda \right\}
\nonumber\\ &&\times\,
\left(\int dx\,{\rm e}^{-i\omega x-F(x)}\right)
\frac{\partial f}{\partial E},
\label{Gnon}
\end{eqnarray}
and
\begin{eqnarray}
\delta G&=&\frac{e^2}{2}\,{\rm Im}\int\frac{dEd\omega}{(2\pi)^2}\,K_\omega[1-2f(E+\omega)]\frac{\partial f(E)}{\partial E}
\nonumber\\ &&\times\,
{\rm tr}\big\{[\hat S^+(E)\hat\Lambda\hat S(E)\hat\Lambda-\hat S^+(E+\omega)\hat\Lambda\hat S(E+\omega)\hat\Lambda]
\nonumber\\ &&\times\,
[1-\hat S^+(E)\hat S(E+\omega)]\big\}.
\label{deltaGint}
\end{eqnarray}
In Eq. (\ref{deltaGint}) we again set ${\cal F}=1.$ 
This is sufficient in the perturbative regime to
be studied below.

\subsection{Perturbation theory and renormalization of the S-matrix}

The above general expressions allow to conveniently proceed
with the perturbation theory in the interaction. In order to obtain the
first order correction to the conductance one should expand $G_F$ (\ref{Gnon})
in $F(x)$ and combine linear in $F(x)$ terms of this expansion with the
interaction correction $\delta G$ (\ref{deltaGint}). Then one finds 
$G=G_L+\delta G_{1}+\delta G_{2}$, where $G=G_L$ is the 
Landauer conductance of the dot in the absence of interactions,
\begin{equation}
G_L=-\frac{e^2}{2}\int\frac{dE}{2\pi}\,{\rm tr}[\hat 1-\hat S^\dagger(E)\hat\Lambda\hat S(E)\hat\Lambda]
\frac{\partial f}{\partial E},
\label{GL}
\end{equation}
and the two corrections $\delta G_{1,2}$ read
\begin{eqnarray}
\delta G_{1}=\frac{e^2}{2}\,{\rm Im}\int \frac{dEd\omega}{(2\pi)^2}K_\omega
\left(\coth\frac{\omega}{2T}-\tanh\frac{E+\omega}{2T}\right)
\nonumber\\ \times\,
{\rm tr}\big\{\hat S^\dagger(E+\omega)\hat\Lambda\hat S(E+\omega)\hat\Lambda-\hat S^\dagger(E)\hat\Lambda\hat S(E)\hat\Lambda\big\}
\frac{\partial f(E)}{\partial E},\;\;\;
\label{inel}
\end{eqnarray}
\begin{eqnarray}
\delta G_{2}&=&
-\frac{e^2}{2}\,{\rm Im}\int\frac{dEd\omega}{(2\pi)^2}\,K_\omega\,\tanh\frac{E+\omega}{2T}\,
\frac{\partial f(E)}{\partial E}
\nonumber\\ &&\times\,
{\rm tr}\big\{\hat S^\dagger(E)\hat\Lambda\hat S(E)\hat\Lambda\hat S^\dagger(E)\hat S(E+\omega)
\nonumber\\ &&
-\,\hat S^\dagger(E)\hat\Lambda\hat S(E+\omega)\hat\Lambda
\big\}.
\label{el}
\end{eqnarray}
We note that in Eq. (\ref{inel}) the terms containing 
$\coth (\omega /2T)$ come from the expansion of (\ref{Gnon}) while 
all the tanh-terms originate from Eq. (\ref{deltaGint}). 

It is easy to see that $\delta G_{1}\equiv 0$ 
for any $\hat S(E)$ and at any temperature. 
Indeed, making a shift $E+\omega\to E$ in the term
involving the product $\hat S^\dagger(E+\omega)\hat\Lambda\hat
S(E+\omega)\hat\Lambda$ one reduces the expression under the integral
to the form ${\rm Im}K_\omega\, {\cal B}(E,\omega),$
where 
\begin{eqnarray}
{\cal B}(E,\omega)&=&\left(\coth\frac{\omega}{2T}-\tanh\frac{E}{2T}\right)
\frac{\partial f(E-\omega)}{\partial E}
\nonumber\\ &&
-\,\left(\coth\frac{\omega}{2T}-\tanh\frac{E+\omega}{2T}\right)
\frac{\partial f(E)}{\partial E}.
\end{eqnarray}
Since $f(E)$ is the Fermi function, one has ${\cal B}(E,\omega)={\cal
  B}(E,-\omega)$. At the same time, as follows from Eq. (\ref{K}),
${\rm Im}K_\omega$ is an odd function of $\omega$.  
Hence, the integral over $\omega$ vanishes identically, the term
$\delta G_1$
drops out and only the combination
(\ref{el}) needs to be taken into consideration. 

Let us make a shift  $\hat S(E)\to \hat S(E)+\delta\hat S(E)$ in
the Landauer formula (\ref{GL}). Then the corresponding linear in 
$\delta\hat S(E)$ correction to the conductance becomes
\begin{eqnarray}
\delta G_L&=&\frac{e^2}{2}\int\frac{dE}{2\pi}\,{\rm tr}[\delta\hat S^\dagger(E)\hat\Lambda\hat S(E)\hat\Lambda
\nonumber\\ &&
+\,\hat S^\dagger(E)\hat\Lambda\delta\hat S(E)\hat\Lambda]\frac{\partial f}{\partial E}.
\label{dGL}
\end{eqnarray} 
Comparing this formula to the first order correction  
(\ref{el}) we can choose $\delta\hat S(E)$ and $\delta\hat S^\dagger(E)$ in the form:
\begin{eqnarray}
\delta \hat S=\frac{1}{2i}\int\frac{d\omega}{2\pi}\,K_\omega
\bigg[\tanh\frac{E+\omega}{2T}[\hat S(E+\omega)-\hat S(E)]
\nonumber\\ 
+\,\tanh\frac{E-\omega}{2T}\,\hat S(E)[\hat S^\dagger(E-\omega)-\hat S^\dagger(E)]\hat S(E)\bigg],
\nonumber\\
\delta \hat S^\dagger=\frac{-1}{2i}\int\frac{d\omega}{2\pi}\,K^*_\omega
\bigg[\tanh\frac{E+\omega}{2T}[\hat S^\dagger(E+\omega)-\hat S^\dagger(E)]
\nonumber\\ 
+\,\tanh\frac{E-\omega}{2T}\,\hat S^\dagger(E)[\hat S(E-\omega)-\hat S(E)]\hat S^\dagger(E)\bigg].
\hspace{0.3cm}
\label{deltaS}
\end{eqnarray}
Although this choice is obviously not a unique one, 
we will stick to it
for the reasons which will become clear below. 
Making use of the property $K_{-\omega}=K^*_{\omega},$ it is easy to check
that the conditions $\big(\delta\hat S^\dagger(E)\big)^\dagger=\delta\hat S(E)$ and 
$\delta \hat S^\dagger(E)\hat S(E)+\hat S^\dagger(E)\delta\hat S(E)=0$ are identically
fulfilled, i.e. the first order renormalization
(\ref{deltaS}) preserves the unitarity of the $S-$matrix. 

In order find the function $K_\omega$ (\ref{K}) for chaotic
quantum dots one needs to perform the average of the product of
two $S-$matrices at different energies (\ref{u}).  Neglecting small localization corrections, 
and applying the formula (\ref{two}) we obtain
\begin{equation}
{\rm tr}\langle\hat 1-\hat S^\dagger(E)\hat S(E+\omega)\rangle=\frac{-i\omega\tau_Dg}{2(1-i\omega\tau_D)}.
\label{T}
\end{equation}
Combining (\ref{T}) with (\ref{K},\ref{u}), we get
\begin{equation}
K_\omega =\frac{e^2}{(-i\omega+0)\left(-i\omega
C_\Sigma+\frac{e^2}{2\pi}g\frac{-i\omega\tau_D}{1-i\omega\tau_D}\right)}
\label{K00}
\end{equation}
and after the the Fourier transformation we arrive at the
expression
\begin{equation}
K(t)=\frac{2\pi}{g}\theta(t)\left(\frac{\tau_D^2(1-{\rm
e}^{-t/\tau})}{(\tau_D+\tau_0)^2} 
+\frac{t}{\tau_0+\tau_D}\right). \label{K000}
\end{equation}
Here we have defined the classical $RC-$time of
the central island
$\tau_0=2\pi C_\Sigma/(e^2g)$   and  $\tau=\tau_D\tau_0/(\tau_D+\tau_0)$.
For large quantum dots one typically has $\tau_0\ll\tau_D$ and, 
hence, $\tau\approx\tau_0.$

Eq. (\ref{K000}) allows to specify the necessary condition of 
applicability for our analysis.  Let us recall that during our 
derivation we have
expanded the effective action in powers of the fluctuating phase
field $\varphi^-.$ The least action condition
fixes this field to be equal to $\varphi^-(t)=K(t)$. Since all the
available time integrals are effectively cut at times exceeding
$\tau_D,$ it is sufficient to require $\varphi^-(\tau_D)=K(\tau_D)\ll
2\pi.$ Then from Eq. (\ref{K000}) we
conclude that our expansion of the exact
effective action in powers of the field $\varphi^-(t)$
is justified in the ``metallic'' limit
\begin{equation}
g \gg 1.
\label{condition}
\end{equation} 
We also note that this condition is necessary but in general not 
a sufficient one in order to justify the perturbative expansion 
in the interaction employed in this subsection. An additional condition will be
established below in Sec. 4C.

Consider the most relevant physical limit $\tau_D\gg\tau_0$. In this
case one has $K_\omega\approx\frac{2\pi}{g}\left(\frac{1}{-i\omega}
-\frac{1}{1/\tau_0-i\omega}\right).$ 
Since the integral in Eqs. (\ref{deltaS})
is taken over a wide range of energies, the main logarithmic 
correction to the scattering matrix can be obtained if one replaces $\hat S(E+\omega)$ by its 
energy independent average $\langle \hat S(E)\rangle=\hat R$.
Then one finds  
\begin{eqnarray}
\delta\hat S(E)&=&\frac{\ln(1/E\tau_0)}{g}\big[\langle\hat S\rangle-\hat S(E)
\nonumber\\ &&
-\,\hat S(E)\big(\langle\hat S^\dagger\rangle-\hat S^\dagger(E)\big)\hat S(E)\big].
\label{ddS}
\end{eqnarray}
Eq. (\ref{ddS}) can be used to derive  renormalization
group (RG) equations for the energy dependent scattering matrix of a
quantum dot. Following the standard procedure let us fix an
infinitesimal energy interval between $E$ and $E-dE$. From Eq. 
(\ref{ddS}) one easily finds the corresponding correction $d\hat S(E)$ 
to the scattering matrix. Lowering the energy and repeating 
this procedure many times one arrives
at the following RG equations
\begin{equation}
\frac{d\hat S(E)}{d\ln(1/E\tau_0)}=\frac{\langle\hat S\rangle
-\hat S(E)\langle\hat S^\dagger\rangle\hat S(E)}{g},
\label{dSdln}
\end{equation}
where the effective
conductance $g$ is expressed via the renormalized scattering
matrix at a given energy by means of the Landauer formula. Making use
of Eqs. (\ref{ST}-\ref{ttt}) we define
\begin{equation}
g={\rm tr}\hat T^\dagger\hat T.
\label{Landfla}
\end{equation}
 Hence, $g$
itself gets renormalized and becomes energy dependent. 
Finally, substituting $\hat S(E)$
in the form (\ref{ST}) we observe that
$\hat U(E)$ remains unchanged in the
course of renormalization.
Thus Eq. (\ref{dSdln}) describes the renormalization
of the barrier transmissions  connecting
the chaotic cavity to the ideal leads.  
After simple manipulations and making use of Eq. (\ref{Landfla})
we arrive at the scaling equations for the
transmission matrices $\hat T^+\hat T$:
\begin{equation}
\frac{d\hat T^\dagger\hat T}{d\ln(E\tau_0)}=
\frac{2\big(\hat 1-\hat T^\dagger\hat T\big)\hat T^\dagger\hat T}{{\rm
    tr}\hat T^\dagger\hat T} .
\label{RGTT}
\end{equation}
Eq. (\ref{RGTT}) coincides with that recently derived in
Ref. \onlinecite{BN} by means of a different approach. 

Let us emphasize again that
the whole analysis of the present subsection is valid within
the first order perturbation theory in the interaction. Hence, the
above RG equation can be applied only as long as the renormalized
conductance $g(E)$ remains large.

\subsection{I-V curve in the voltage-biased limit}

Provided temperature and/or voltage exceed the level spacing in
the dot and provided one is not interested in resolving subtle
details of the $I-V$ curve related to mesoscopic fluctuations, it
is convenient to perform averaging of the above general
expressions for the current over such fluctuations. The average
values of the products of reflection and transmission matrices
depend only on $\omega,$ not on $E.$ Hence, one can 
integrate over $E$ and derive the current-voltage characteristics
of a quantum dot in the presence of interactions. In particular,
the term $I_0$ in Eq. (\ref{I01}) acquires the standard Landauer
form
\begin{equation}
I_0=\frac{e^2}{\pi}{\rm tr}\langle\hat t^\dagger(E)\hat t(E)\rangle
V=G_0V. \label{I0av}
\end{equation}
In the case of chaotic quantum dots one finds \cite{BB}:
\begin{eqnarray}
{\rm tr}\langle\hat t^\dagger(E)\hat t(E)\rangle=\frac{g_Lg_R}{2g}
\hspace{3cm}
\nonumber\\ 
+\,\nu \frac{g_Lg_R^2(1-\beta_L)+g_L^2g_R(1-\beta_R)}{g^3},
\label{Land}
\end{eqnarray}
where $\nu=-1,$ 0  or 1/2 respectively for
circular orthogonal, unitary and symplectic ensembles. 
 The second term in Eq. (\ref{Land}) represents 
the weak localization (WL) correction. For the model under consideration this
correction is parametrically ($\sim 1/g$) small as compared to the
first term in (\ref{Land}). Below we will concentrate
on the interaction correction to the current $\delta I,$
which is of the same order in $1/g$, but 
nevertheless strongly exceeds the WL correction at sufficiently 
low energies. 

Let us consider the perturbative in the interaction regime.
After averaging over mesoscopic fluctuations the interaction 
correction (\ref{deltaI}) takes the form
\begin{equation}
\delta I=-\frac{e}{\pi}\,\int\frac{d\omega}{2\pi}\,{\rm
Im}\big(K_\omega D (\omega)\big)\,
(\omega-eV)\coth\frac{\omega-eV}{2T}, \label{ILRav}
\end{equation}
where we defined
\begin{eqnarray}
D(\omega)&=&{\rm tr}\langle\big[\hat r^+(E+\omega)\hat
  t'(E+\omega)-\hat r^+(E)\hat t'(E)\big]
\nonumber\\&&
\times\,\big[\hat{t'}^+(E)\hat r(E+\omega)+\hat{r'}^+(E)\hat
  t(E+\omega)\big]\rangle.
\label{Dij}
\end{eqnarray}
All other contributions to the interaction correction vanish upon averaging.
An explicit evaluation of the function $D(\omega )$ (\ref{Dij}) is easily performed
with the aid of Eq. (\ref{W}). Here we only quote the result:
\begin{eqnarray}
D(\omega)=\frac{g_Lg_R(g_R\beta_L+g_L\beta_R)}{2(g_L+g_R)^2}\,
\frac{\omega^2\tau_D^2}{(1-i\omega\tau_D)^2}.
\label{DLR}
\end{eqnarray}
Combining Eqs. (\ref{ILRav}), (\ref{K00}) and
(\ref{DLR}), we obtain
\begin{eqnarray}
\delta I =  -\frac{eB}{2} \int\frac{d\omega}{2\pi }\,
\frac{1}{\omega}\bigg(\frac{1}{1+\omega^2\tau^2}-\frac{1}{1+\omega^2\tau_D^2}\bigg)
\hspace{1cm}
\nonumber\\ \times
\bigg[(\omega+eV)\coth\frac{\omega+eV}{2T}-(\omega-eV)\coth\frac{\omega-eV}{2T}\bigg],
\label{curr}
\end{eqnarray}
where we defined
\begin{equation}
B=g_Lg_R(g_L\beta_R+g_R\beta_L)/(g_L+g_R)^3.
\label{BBB}
\end{equation}
One can also rewrite Eq. (\ref{curr}) in the form
\begin{equation}
\delta I=-\frac{eB}{\pi} \int_0^\infty dt\,\frac{\pi^2
T^2}{\sinh^2\pi Tt} {\rm e}^{-t/\tau_D}(1-{\rm
e}^{-t/\tau_0})\sin eVt. \label{dI}
\end{equation}
The integral can be evaluated analytically. We find
\begin{eqnarray}
\delta I=-\frac{eB}{\pi} {\rm
Im}\bigg[\left(\frac{1}{\tau}+ieV\right)\Psi\left(1+\frac{1}{2\pi T\tau}+i\frac{eV}{2\pi T}\right)
\nonumber\\
-\left(\frac{1}{\tau_D}+ieV\right)\Psi\left(1+\frac{1}{2\pi T\tau_D}+i\frac{eV}{2\pi T}\right)\bigg], 
\label{dIan}
\end{eqnarray}
where $\Psi (x)$ is the digamma 
function. This is a complete
expression for the first order interaction correction to the current in
chaotic quantum dots with large conductances in the voltage-biased regime.

We will now analyze the above general expression in various
specific limits. At $T\to 0$ the $I-V$ curve
reduces to the following simple form
\begin{equation}
\frac{dI}{dV}=G_0-\frac{B}{R_q}
\ln\frac{\tau_D^2(1+(eV\tau)^2)}{\tau^2(1+(eV\tau_D)^2)}.
\end{equation}
Provided the voltage is large $eV\tau \gg 1$ the interaction
correction turns out to be small. In the limit $\tau_0 \ll \tau_D$
one finds
\begin{equation}
\frac{dI}{dV}=G_0-\frac{Bg^2E_C^2}{\pi R_qe^2V^2},
\label{largeV}
\end{equation}
where $E_C=e^2/2C_\Sigma$ is the charging energy.
In the intermediate range of voltages $1/\tau_D \ll eV \ll 1/\tau_0$
the interaction correction becomes logarithmic in $V$
\begin{equation}
\frac{dI}{dV}=G_0+\frac{2B}{R_q} \ln\left(eV\tau_0 \right)
\label{logV}
\end{equation}
and at lower voltages and temperatures $eV,T \ll 1/\tau_D$ we obtain
\begin{equation}
\frac{dI}{dV}=G_0-\frac{2B}{R_q}\left[\ln\frac{\tau_D}{\tau}-
\left(\frac{(eV)^2}{2}+\frac{\pi^2T^2}{3}\right)(\tau_D^2-\tau^2)\right].
\label{sat}
\end{equation}
The latter expression demonstrates that in the regime
under consideration the linear conductance of the quantum dot remains
non-zero down to zero temperature.

Taking the limit $V \to 0$ in the
general results (\ref{dI},\ref{dIan}) one can derive the linear
conductance of the dot at arbitrary temperatures. Evaluating the integrals, we
obtain
\begin{eqnarray}
G&=& G_0 -\frac{2B}{R_q} \left[L(T\tau)- L(T\tau_D)\right],
\label{dGan}
\end{eqnarray}
where we defined 
\begin{equation}
L(x)=\Psi \left(1+\frac{1}{2\pi x}\right)+\gamma+
\frac{1}{2\pi x}\Psi' \left(1+\frac{1}{2\pi x}\right)
\label{L}
\end{equation}
and $\gamma \simeq 0.577$ is the Euler constant.
As before, this general expression can be simplified further in
various limits. At high temperatures $T \gg gE_C$ we find
\begin{equation}
G=G_0-\frac{Bg}{R_q}\left\{\frac{E_C}{3T}-
\frac{3\zeta(3)g}{2\pi^4} 
 \left(\frac{E_C}{T}\right)^2\right\},
\label{3T}
\end{equation}
where $\zeta(3)\simeq 1.202$.  
In the regime $1/\tau_D \ll T \ll 1/\tau_0$ we again arrive at the 
logarithmic interaction correction to the conductance 
\begin{equation}
G=G_0+\frac{2B}{R_q} \ln\left(T\tau_0\right)
\label{logT}
\end{equation}
which crosses over to Eq. (\ref{sat}) as $T$ becomes
smaller than the inverse dwell time $1/\tau_D$.

The above results completely describe perturbative interaction
correction to the current-voltage characteristics of highly conducting
chaotic quantum dots in the voltage-biased regime. We also note that in 
certain limits these results reduce to
ones obtained previously \cite{GZ00,BN,ABG,SR} by means of different
techniques. For instance, the logarithmic behavior of the interaction
correction (\ref{logT}) can easily be recovered
from the RG approach \cite{BN} which we also discussed in Sec. 4B.  On the 
other hand, some other
results, e.g., Eqs. (\ref{largeV},\ref{3T}), cannot be obtained within 
this RG scheme.

Under which conditions do the above perturbative results remain valid?
At high temperatures the validity of the result (\ref{3T}) 
is guaranteed by the inequality $T \gg gE_C$. In this limit no
additional requirement (\ref{condition}) is needed,
i.e. Eq. (\ref{3T}) applies also for $g \sim 1$. The same is true for
the result (\ref{largeV}) derived in the limit of large voltages. 
On the other hand, at lower temperatures and voltages $T,eV \ll gE_C$
the condition (\ref{condition}) should be satisfied
ensuring that fluctuations of the phase $\varphi^-(t)$
remain weak.

An additional applicability condition for the above results 
is obtained if one requires fluctuations of the phase 
$\varphi^+(t)$ to be small. Let us recall that in the course of our
derivation we have approximated
the function ${\cal F}$ (\ref{calF}) by unity. This approximation 
is appropriate as soon as the function $F(t)=\langle
(\varphi^+(t)-\varphi^+(0))^2\rangle /2$ remains much smaller than
one and fluctuations of the phase $\varphi^+(t)$
can be disregarded. Hence, the condition 
\begin{equation}
F(\tau_F) \sim 1 
\label{tF}
\end{equation}
defines yet one more time scale in our problem, $\tau_F$, which restricts the
applicability of the above perturbative results. 

In order to determine the time scale $\tau_F$ it suffices to 
evaluate the function $F(t)$ in the limit of low voltages $V \to 0$.
In this limit Eq. (\ref{Ffull}) reduces to a simple expression
\begin{equation}
F(t)=\int\frac{d\omega}{2\pi}\, {\rm Im} (K_\omega )
\coth\frac{\omega}{2T} (1-\cos\omega t).
\label{FV0}
\end{equation}
Combining this expression with  Eq. (\ref{K00}) one finds
 \begin{equation}
F(t)=
\frac{\pi Tt^2}{g(\tau_0+\tau_D)}+
\frac{2\kappa}{g}
\int_0^\infty d\omega\,\frac{\omega\coth\frac{\omega}{2T}}{1+\omega^2\tau^2}\,
\frac{1-\cos\omega t}{\omega^2},
\label{Fexp}
\end{equation}
where $\kappa=\tau_D^2/(\tau_D+\tau_0)^2.$
In the
most interesting limit $T \to 0$ the integral in Eq. (\ref{Fexp}) yields within the logarithmic
accuracy $F(t)\simeq (2\kappa /g)
\ln (t/\tau)$. Then from Eq. (\ref{tF}) we obtain
\begin{equation}
\tau_F \sim \tau \exp (g/2\kappa).
\label{exppp}
\end{equation}
Thus fluctuations of the phase $\varphi^+(t)$ can be neglected at all
temperatures and voltages only provided the dwell time $\tau_D$ is
much smaller than the parameter $\tau_F$ (\ref{exppp}). More
generally, the results presented in this
subsection are correct under the condition
\begin{equation}
{\rm max}(T, eV, \tau_D^{-1}) \gg \tau_F^{-1}.
\label{cond2}
\end{equation}
It is easy to check that in this case the interaction correction 
in Eqs. (\ref{logV}), (\ref{sat})
and (\ref{logT}) remains small as compared to $G_0$.
If the condition (\ref{cond2}) is violated the system enters an
essentially non-perturbative
regime in which case more accurate analysis becomes necessary. This
analysis is beyond the scope of the present paper. Let us also
point out that, while the non-perturbative regime
is important from a theoretical point of view, it does not
seem to be of much practical relevance for  
quantum dots with $g \gg 1$ considered here.

\subsection{Effect of external impedance}
So far all our final results have been formulated for 
the case of ideally conducting external leads $Z_S(\omega)\to 0$. 
In many experimental situations, however, this voltage-biased model is
not appropriate since the impedance of external leads remains non-zero 
in the relevant frequency range. 
For this reason it is important to find out how the above results
for the $I-V$ curve are modified in the case $Z_S(\omega)\not=0$.

In order to answer this question we will make use of the
general expressions for the current derived in Appendix D.
Averaging these expressions over
mesoscopic fluctuations we arrive at the result
\begin{equation}
I(V)=I_0(V)+\delta I_{\rm tot}(V),
\end{equation}
where $I_0(V)$ remains the same and is given by Eq. (\ref{I0av}),
while $\delta I_{\rm tot} (V)$ represents the total interaction
correction which takes the form:
\begin{eqnarray}
\delta I_{\rm tot}(V)=-\frac{e}{2\pi}\,\int\frac{d\omega}{2\pi}\,
\bigg((\omega+eV)\coth\frac{\omega+eV}{2T}
\hspace{0.5cm}
\nonumber\\
-\,(\omega-eV)\coth\frac{\omega-eV}{2T}\bigg)
{\rm Im}\sum_{ij}K_{ij}(\omega)D_{ji}(\omega),
\label{ILRavimp}
\end{eqnarray}
where
\begin{eqnarray}
D_{RL}(\omega)&=&{\rm tr}\langle\hat t'(E+\omega)\hat{t'}^\dagger(E)\hat r(E+\omega)\hat r^\dagger(E+\omega)\rangle,
\nonumber\\
D_{LL}(\omega)&=&{\rm tr}\langle\hat r^\dagger(E)\hat t'(E)\hat{t'}^\dagger(E)\hat r(E+\omega)\rangle,
\nonumber\\
D_{RR}(\omega)&=&{\rm tr}\langle\hat t(E+\omega)\hat t^\dagger(E+\omega)\hat r'(E+\omega)\hat{r'}^\dagger(E)\rangle,
\nonumber\\
D_{LR}(\omega)&=&{\rm tr}\langle\hat t(E+\omega)\hat t^\dagger(E)\hat r'(E)\hat{r'}^\dagger(E)\rangle.
\end{eqnarray}
The above expressions are still rather cumbersome. For the sake
of simplicity below we will analyze the case of symmetric 
quantum dots  with $g_L=g_R=g/2,$
$\beta_L=\beta_R=\beta,$ $R_L=R_R=R$ and $C_L=C_R=C.$ We will also assume
the external impedance to be Ohmic $Z_S(\omega)=R_S$ at all relevant 
frequencies. In this case by virtue of Eq. (\ref{W}) we find
\begin{eqnarray}
D_{LL}=D_{RR}=\frac{g\beta}{16}
+\frac{g(1-2\beta)}{32(1-i\omega\tau_D)}
+\frac{g\beta}{32(1-i\omega\tau_D)^2},
\nonumber\\
D_{LR}=D_{RL}=\frac{g(1+2\beta)}{32(1-i\omega\tau_D)}
-\frac{g\beta}{32(1-i\omega\tau_D)^2}.
\hspace{0.5cm}
\label{Dij11}
\end{eqnarray}
From Eqs. (\ref{Kij}, \ref{calA}) we also obtain $K_{LL}=K_{RR}=K_S+K,$
$K_{LR}=K_{RL}=K_S-K,$ where
\begin{eqnarray}
K&=&\frac{e^2}{(-i\omega+0)\left(-2i\omega C+\frac{-i\omega C_g}{1-i\omega R_SC_g/4}
+\frac{2}{R}\frac{-i\omega\tau_D}{1-i\omega\tau_D}\right)},
\nonumber\\
K_S&=&\frac{e^2}{(-i\omega+0)(-2i\omega C+\frac{4}{R_S}+\frac{2}{R})}.
\label{KRS}
\end{eqnarray}
To simplify the analysis further let us 
replace $1-i\omega R_SC_g/4$ by $1$ in the denominator
of the expression for $K$ (\ref{KRS}). This is appropriate either
for $C_g \to 0$ or in the limit $R_S\ll 2R$. Then we get
\begin{equation}
\delta I_{\rm tot}(V)=\delta I(V) +\delta I_S(V),
\end{equation}
where $\delta I(V)$ is the interaction correction evaluated 
for $Z_S=0$ and $\delta I_S(V)$ represents an additional
contribution due to the external shunt. 
The term $\delta I(V)$ is defined by 
Eqs. (\ref{curr},\ref{dI},\ref{dIan}), whereas for $\delta I_S(V)$
one finds
\begin{eqnarray}
\delta I_S(V)=-\frac{eg}{16\pi}\,\int\frac{d\omega}{2\pi}\,
{\rm Im}\left(K_S(\omega)\left(\beta+\frac{1}{1-i\omega\tau_D}\right)
\right)
\nonumber\\ \times
\bigg[(\omega+eV)\coth\frac{\omega+eV}{2T}-(\omega-eV)\coth\frac{\omega-eV}{2T}\bigg].
\hspace{0.5cm}
\end{eqnarray}
This formula can also be transformed to the following expression
\begin{eqnarray}
\delta I_S(V)
=-\frac{e}{4\pi}\frac{g}{g+4g_S}\int_0^\infty dt\frac{\pi^2 T^2}{\sinh^2\pi Tt}\sin eVt
\nonumber\\
\times\,\bigg[\beta(1-{\rm e}^{-t/\tau_S})
+1-\frac{\tau_D{\rm e}^{-t/\tau_D}-\tau_S{\rm e}^{-t/\tau_S}}{\tau_D-\tau_S}\bigg],
\label{IS}
\end{eqnarray}
where we defined $\tau_S=RR_SC/(R_S+2R)$. As before, this integral can
be expressed in terms of the $\Psi$-functions. For the sake of brevity
we will omit the corresponding expressions here. 

Let us concentrate on the linear in voltage regime.
Making use of the above general results one can easily determine
the total linear conductance $G$ of the dot in the presence of an external
Ohmic shunt. It reads
\begin{eqnarray}
G= G_0 -\frac{1}{2R_q}\bigg\{ \beta \big(L(T\tau)- L(T\tau_D)\big)
+\frac{g}{g+4g_S}
\nonumber\\ 
\times \bigg[\frac{\tau_D\big(L(T\tau_D)-L(T\tau_S)\big)}{\tau_D-\tau_S}
+(\beta+1)L(T\tau_S)
\bigg]\bigg\},
\label{dGS}
\end{eqnarray}
where the function $L(x)$ was defined in Eq. (\ref{L}).

For large quantum dots one typically has $\tau_D\gg\tau_0>\tau_S.$
In this case at sufficiently low temperatures $ T\tau_D\ll 1$ 
the whole quantum dot can be considered
as a single scatterer, and we find
\begin{eqnarray}
G&=&\tilde G_0
-\frac{2\tilde \beta}{R_q}\frac{g}{g+4g_S}
\left(\gamma+1+\ln\frac{1}{2\pi T\tau_D}\right),
\label{llT}
\end{eqnarray}
where
\begin{eqnarray}
\tilde G_0=G_0-\frac{\beta }{2R_q}\left(\ln\frac{\tau_D}{\tau}+
\frac{g}{g+4g_S}\ln\frac{\tau_D}{\tau_S}\right)
\label{reng0}
\end{eqnarray}
and
$\tilde\beta=(\beta+1)/4$
is the effective total Fano factor of a symmetric chaotic quantum dot, 
see also Eq. (\ref{Fano}).
Eqs. (\ref{llT}), (\ref{reng0}) smoothly match with
the result \cite{GZ00} derived for a single coherent
scatterer with $\tau_D \lesssim \tau_0$. 

In the opposite high temperature limit $T\tau_D\gg 1$, albeit $T\tau_0\ll 1,$
we obtain
\begin{eqnarray}
G&\approx& G_0 - \frac{\beta}{2R_q}\frac{g}{g+4g_S}\ln\frac{\tau_0}{\tau_S}
\nonumber\\ &&
-\,\frac{\beta}{R_q}\frac{g+2g_S}{g+4g_S}\left(\gamma+1+\ln\frac{1}{2\pi T\tau_0}\right).
\label{hhT}
\end{eqnarray}

\begin{figure}
\includegraphics[width=8.5cm]{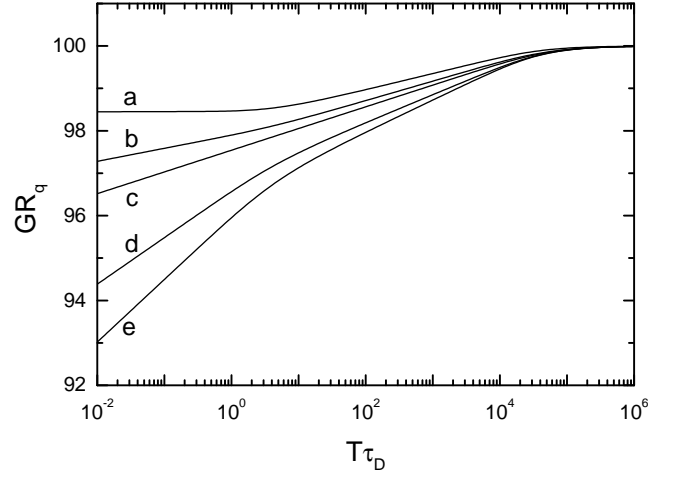}
\caption{Conductance of a symmetric quantum dot, as given by Eq.
(\ref{dGS}). The Fano factor of each barrier 
is chosen to be $\beta =1/3,$ its conductance is $200/R_q$
(hence $G_0=100/R_q$)
and $\tau_D/\tau_0=10^{4}.$ Different curves correspond to different
values of the shunt resistor: (a) $R_S=2.5\times 10^{-5}\, R_q$, (b)
$R_S=0.0025\, R_q,$
(c) $R_S=0.005\, R_q,$ (d) $R_S=0.025 R_q,$ and (e) $R_S=0.25 R_q.$ A
crossover between the two logarithmic regimes is clearly observed 
at $T\tau_D \sim 1$.}
\label{cond111}
\end{figure}

Let us compare the results (\ref{llT}) and (\ref{hhT}). In both
temperature regimes we observe that the interaction correction depends
logarithmically on temperature, though with different prefactors in
front of the logarithm. At high temperatures $T\tau_D \gg 1$ this
prefactor is determined by a sum of two different contributions originating
from the terms $\delta I$ and  $\delta I_S$. In this regime the two
barriers of the quantum dot behave as independent ones. In the other 
regime $T\tau_D < 1$ 
the first logarithmic term saturates and does not depend on $T$
anymore providing effective renormalization of the non-interacting 
conductance (\ref{reng0}).
On the contrary, the logarithmic temperature dependence (\ref{llT}) in
the second term survives down to exponentially small temperatures. In
this regime the quantum dot behaves as a single coherent scatterer.
Its internal structure becomes insignificant in this case and the
interaction correction scales with the total Fano factor $\tilde \beta$.
A crossover between the two logarithmic regimes is clearly observed in
Fig. 2. 
It is also important to emphasize that, while the logarithmic in 
$T$ correction (\ref{llT}) vanishes for $R_S \to 0$,
in the opposite limit $R_S\gg R$ it becomes practically independent
of the shunt resistance. 

The presence of two different logarithmic regimes can also be understood
bearing in mind a close relation between shot noise and interaction
effects in mesoscopic conductors. At high temperatures $T\tau_D \gg
1$, i.e. in the regime of independent barriers,
the noise at each of them determines the corresponding
contribution to the interaction correction. Hence, the latter is
proportional to the parameter $\beta$. In the opposite limit $T\tau_D \ll
1$ the barriers are not anymore independent. In this regime the
temperature dependent part of the interaction correction should be
related to the total shot noise of the quantum dot which, as we discussed
in Sec. 3, depends on the parameter $\tilde \beta$. The high frequency 
noise $\omega \tau_D \gtrsim 1$ now only provides the renormalization
of $G_0$ which is again proportional to $\beta$. 

We will continue the discussion of the above results and their
experimental relevance in the next section.

\section{Discussion and conclusions}

The analysis presented here provides a general theoretical description
of electron transport through disordered interacting quantum dots in 
the ``metallic'' limit of large conductances. Our formalism 
combines real time path-integral-based influence functional technique with
the scattering matrix approach. The main technical result of the
present work is the derivation of the effective action for quantum dots
described by {\it arbitrary} energy dependent scattering matrices.
Due to this particular feature our technique allows to investigate
quantum transport of interacting electrons in a very wide class of conductors
which includes, e.g., structures with resonant transmission and many
others. This formalism is a direct generalization of our earlier
approach \cite{GZ00,GGZ02} which embraced structures described by energy 
independent scattering matrices and did not account for
internal dynamics of disordered conductors.

Although our analysis is mainly aimed at interaction effects, the first
and immediate result of our derivation is the expression for
frequency-dependent current noise in non-interacting quantum dots.
In fully symmetric dots the shot noise spectrum turns out
to be defined by the standard expression (\ref{Khlus}) with the total
Fano factor $\tilde\beta =(\beta +1)/4$. For asymmetric quantum dots the
corresponding expression becomes far more complicated.

Turning to interaction effects, let us briefly summarize
our main results for the linear conductance $G$ of 
chaotic quantum dots (characterized
by dimensionless conductance $g\gg 1$ and charging energy $E_C$) connected
to the voltage source via an Ohmic resistor $R_S$. The conductance
$G$ is expressed in a general form 
$$
G=G_0+\delta G(T),
$$ 
where $G_0$
is the conductance of a non-interacting dot and $\delta G$ is
the interaction correction. This correction is negative, i.e. electron-electron
interactions tend to suppress the conductance of quantum dots. 

For the situation
under consideration one typically has $\tau_D \gg \tau_0\sim 1/gE_C>\tau_S
=R_SC/2(1+G_0R_S)$. For simplicity
we quote the results for symmetric quantum dots. At 
high temperatures $T\tau_0 \gg 1$ in the leading
approximation one finds
\begin{equation}
\delta G/G_0\simeq -\frac{\beta \chi E_C}{3T},
\label{12T}
\end{equation}
where $\beta$ is the Fano factor of a single barrier (\ref{beta}) 
and the parameter $\chi$ is defined below.
At lower temperatures $T\tau_0 \ll 1$ the power law dependence 
(\ref{12T}) crosses over into the logarithmic one
\begin{equation}
\delta G\simeq \frac{\beta \chi}{2R_q}\ln (T\tau_0).
\label{chilog}
\end{equation}
The parameter $\chi$ in Eqs. (\ref{12T}), (\ref{chilog}) 
depends on the relation
between $R_S$ and the dot resistance $R_d\equiv 1/G_0$. 
For $R_S \to 0$ one has $\chi =1$,
while in the opposite limit $R_S \gg R_d$ this parameter is 
$\chi =2$. Thus, in both limits  (\ref{12T}) and (\ref{chilog}) the magnitudes
of the interaction correction $\delta G$ evaluated in the
current- and voltage-biased regimes differ by the factor 2. The magnitude
of $\delta G$ is smaller in the latter regime because voltage
fluctuations across the dot are suppressed in this case.

Eq. (\ref{chilog}) applies down to temperatures $T \sim 1/\tau_D$. 
At lower temperatures another logarithmic regime sets in and the
difference between current- and voltage-biased situations becomes more
dramatic. In the limit $T\tau_D \ll 1$ one finds
\begin{equation}
G\simeq \tilde G_0 +\frac{2\tilde \beta}{R_q}\frac{1}{1+4R_d/R_S}\ln (T\tau_D),
\label{lreg}
\end{equation}
where
\begin{equation}
\tilde G_0\simeq G_0-\frac{\beta \chi}{2R_q}\ln \frac{\tau_D}{\tau_0}.
\label{perenorm}
\end{equation}
 Eq. (\ref{perenorm}) describes an effective renormalization
of $G_0$ by electron-electron interactions. This renormalization is again 
twice as big in the current-biased regime as it is
in the voltage-biased one. More importantly, as follows from
Eq. (\ref{lreg}), for any non-zero $R_S$ {\it no saturation} of the 
the logarithmic dependence of the interaction correction on
temperature is expected at $T\tau_D \lesssim 1$. We also observe that at 
$T\tau_D \gtrsim 1$ the interaction
correction $\delta G$ always scales with $\beta$ and, hence, 
vanishes completely for fully transparent 
barriers $\beta \to 0$. By contrast, at lower temperatures 
$T\tau_D \lesssim 1$ only the $T$-independent renormalization of
$G_0$ disappears in this limit, while 
the correction (\ref{lreg}) scales with $\tilde \beta $ and remains finite even
for $\beta =0$. This fact illustrates a direct relation
between shot noise and interaction effects in mesoscopic conductors,
cf. Eqs. (\ref{Khlus}) and (\ref{lreg}). 
We also observe that 
in the limit $\beta \to 0$ the conductance (\ref{lreg}) becomes
completely independent of the charging energy $E_C$.

Similar results can be obtained for a non-linear $I-V$ characteristics
of our system. In particular, at $T \to 0$ the corresponding
expressions are reproduced if one substitutes 
$G \to dI/dV$ and $T \to eV$ in Eqs. (\ref{12T}-\ref{perenorm}).

Both logarithmic regimes in the dependence of the interaction correction on
temperature and voltage discussed above were observed in many
experiments on various mesoscopic structures. Here we will
briefly discuss only a few examples. In the experiment \cite{Weber}
this dependence was found in short diffusive metallic bridges
at temperatures (voltages) $T \lesssim 1$K. No saturation
was observed down to the lowest temperature $< 100$ mK.
Estimating the Thouless
energy $\sim 1/\tau_D$ for the parameters of Ref. \onlinecite{Weber}
one arrives at the value of order few Kelvin. Hence, this
experiment was performed in the regime $T\tau_D < 1$ (\ref{lreg}).

Recently it was argued \cite{SR} that in order to explain
the experimental results \cite{Weber} it is necessary to assume
the presence of insulating barriers at interfaces
between the bridge and the reservoirs. We note, however, that 
this assumption can hardly be justified because for the samples
\cite{Weber} it would inevitably lead to values of $g$
$10\div 100$ times smaller than actually measured. The presence
of insulating barriers can also be ruled out on the experimental
grounds \cite{PC}. This controversy is eliminated if one
recalls that the calculation \cite{SR} was performed within the
voltage-biased model $R_S \to 0$. For this reason the authors
\cite{SR} have overlooked the logarithmic dependence (\ref{lreg})
and attributed the observations \cite{Weber} to the regime $T\tau_D
\gg 1$, hence, requiring much smaller values of $1/\tau_D < 100$ mK.
Since the voltage-biased model does not strictly apply to the
experiments \cite{Weber}, no such requirement is actually needed.

Note, that in some other experiments, e.g., ones with multiwalled
carbon nanotubes \cite{Leuven,Basel,Liu}, the voltage-biased model
appears to be applicable. Both the logarithmic temperature 
dependence \cite{FN4} of $\delta G$ at larger $T$ and its 
saturation at temperatures of order few Kelvin were clearly observed.  
The latter temperature range is consistent with the estimates of 
the parameter $1/\tau_D$ for the nanotubes \cite{Leuven,Basel,Liu}. 

More recently, the logarithmic temperature dependence of the
conductance was observed in strongly disordered 
multiwalled carbon nanotubes \cite{Tark,PC2}. No saturation of this
dependence was found down to the lowest measurement temperature $T
\sim 100$ mK. Although estimates
of the parameter $1/\tau_D$ for the samples \cite{Tark,PC2} 
yield values below 100 mK, it is not easy to interpret
these observations in terms of the high temperature logarithmic regime
(\ref{chilog}) because $L_T \sim \sqrt{D/T}$ ($D$ is the diffusion
coefficient) remains shorter than the length of the nanotube $L$
at all relevant temperatures $T \gtrsim 0.1$ K. In this situation 
our zero-dimensional description cannot be applied to the nanotube
as a whole.

An alternative explanation is based on viewing the 
nanotube as a chain of $N$ connected in series
statistically independent segments of the length 
$L_{\rm s} \approx L/N$. For $L_T \gtrsim L_{\rm s}$
each of the $N$ segments should behave as a zero-dimensional
coherent scatterer shunted by an external impedance effectively
produced by the remaining $N-1$ segments. This scenario applies provided
the scale $L_{\rm s}$ exceeds an effective transversal dimension of the
system $L_{\rm tr}$. In the opposite case $L_{\rm s}< L_{\rm tr}$ a
slightly more general picture of an array of $M\times N$ independent
scatterers can be introduced, where the number $M$ depends on the
ratio $L_{\rm tr}/L_{\rm s}$. Applying the result (\ref{lreg}) for 
each of the segments and assuming $N \gg 1$, at temperatures 
$T\tau_D \lesssim 1$ one finds the conductance of the system in the
form \cite{FN3}
\begin{equation}
G(T)\approx \tilde G_0+\frac{2\tilde\beta}{R_q}\frac{M}{N}\ln (T\tau_D),
\label{ylog}
\end{equation}
where  $\tilde G_0$ represents its conductance at $T\sim 1/\tau_D$,
$M=1$ for $L_{\rm s}> L_{\rm tr}$ and $M>1$ otherwise.
For diffusive nanotubes one has $\tilde \beta \approx 1/3$ and 
$\tau_D \sim L_{\rm s}^2/D$. The crucial point of this scenario is that
at sufficiently low $T$ the scale $L_{\rm s}$ does not depend on temperature
and is set by interactions.

Note that a similar effect was also observed in
disordered metallic wires \cite{MJW}. Clear deviations
from the standard 1d behavior of the interaction correction 
$\delta G \propto 1/\sqrt{T}$ were reported at temperatures 
$T\lesssim 1/\tau_{\varphi}$, where $\tau_{\varphi}$ is the low
temperature dephasing time measured in the same experiment. 
For the parameters \cite{MJW}
one can verify that both the observed magnitude and temperature
dependence of the interaction correction are consistent with
Eq. (\ref{ylog}), where
$L_{\rm s} \sim L_{\varphi}=\sqrt{D\tau_{\varphi}}$ and $M=1$.

Turning again to the data \cite{PC2} let us recall that in this case
the observed values of $L_{\varphi}$ were considerably smaller than
the nanotube circumference $\pi d$. Therefore we can assume 
$L_{\rm s} <\pi d$ and define \cite{FN7} $M\sim \pi d/L_{\rm s}$. 
Then we obtain $M/N \simeq \pi d/L,$   
i.e. $L_{\rm s}$ drops out and for $\tilde \beta =1/3$ the prefactor
in front of 
the logarithmic term in Eq. (\ref{ylog}) becomes $\simeq 2.1 d/R_qL$.
This universal value fits well with the observations \cite{PC2} at $T
\lesssim 10$ K. Finally, we note that the latter inequality is
also consistent with the condition $T\tau_{\varphi} \lesssim 1$ because 
the measured values of $\tau_{\varphi}$ were found to be in the range
$\tau_{\varphi} \sim 10^{-12}$ sek. Hence,
the assumption $L_{\rm s} \sim L_{\varphi}$ and $\tau_D \sim \tau_{\varphi}$ 
appears to be supported by the data \cite{PC2} as well.

Further experimental and theoretical investigations
of the effect of interactions on low temperature electron transport 
in disordered conductors are warranted.

 We would like to thank D.A. Bagrets, A.V. Galaktionov, Yu.V. Nazarov
and S.V. Sharov for interesting discussions and P. Hakonen, P. Mohanty,
M. Paalanen and H. Weber for patiently explaining us various details of their
data and experiments \cite{Weber,MJW,Tark}. 
This work is part of the Kompetenznetz ``Funktionelle Nanostructuren''
supported by the Landestiftung Baden-W\"urttemberg gGmbH. 

\appendix
\section{Evolution operator}

For non-interacting systems the relation between the $S-$matrix and 
the Green functions, or, which is the same, the evolution operator
is known since the work by Fisher and Lee \cite{FL}.
Here we use the inverse form of this relation, namely we express
the evolution operator in terms of the $S-$matrix.
We introduce the components of the evolution operator
$U_{ij}(t,0;r,r'),$ where $i,j$ label 
the left ($l$) and the right ($r$) leads as well as the dot ($d$).
The component $U_{ll}$ reads
\begin{eqnarray}
U_{ll}^{\alpha\beta}(t,0;r,r')
&=&\frac{1}{\sqrt{v_\alpha^lv_\beta^l}}\int\frac{dE}{2\pi}\,e^{-iEt+iE(r/v_\alpha^l-r'/v_\beta^l)}
\nonumber\\&& \times\,
\big[\delta_{\alpha\beta}+\theta(r)\theta(-r')(r_{\alpha\beta}(E)-\delta_{\alpha\beta})
\nonumber\\ &&
+\,\theta(-r)\theta(r')(r^+_{\alpha\beta}(E)-\delta_{\alpha\beta})\big].
\nonumber
\end{eqnarray}
Here $\alpha,\beta$ are the channel indices, $v_\alpha^j$ is the
velocity in the $\alpha-$th channel
of $j-$th lead.
Three other components of the evolution operator $U_{rr},$ $U_{lr}$ and $U_{rl}$ are expressed analogously.
It is convenient to introduce the ``flight times'' in the 
$\alpha-$th channel $x = r/v_\alpha^j,$ $y = r'/v_\alpha^j$
and express the wave functions and the evolution operator via 
these new variables.
The velocities $v_\alpha^j$ then disappear, and we can conveniently rewrite
the above relation in the matrix form
\begin{eqnarray}
\hat U(t,0;x,y)=\hat 1\delta(t-x+y)
\hspace{3.3cm}
\nonumber\\
+\,\theta(x)\theta(-y)[\hat S(t-x+y)-\hat 1\delta(t-x+y)]
\hspace{0.65cm}
\nonumber\\
+\,\theta(-x)\theta(y)[\hat S^+(-t+x-y)-\hat 1\delta(t-x+y)].
\label{U0}
\end{eqnarray}
Note that here we present $\hat U$ in the form of a $2\times 2$ 
(rather than $3\times 3$) matrix leaving out the components
describing transitions (e.g. from the dot to the dot
and some others) which turn out to be unimportant for
our analysis. We will return to this point in Appendix B.

The operator (\ref{U0}) has the following properties:
(i) $\hat U^\dagger(t,0)=\hat U(0,t),$ and (ii)
$\lim\limits_{t\to +\infty}\hat U^\dagger(t,0)\hat U(t,0)=\hat 1.$
At any finite time $t$ this operator does not obey the unitarity
condition. This is quite natural because the components 
describing the evolution
in the dot are missing in Eq. (\ref{U0}).   However
the operator
$\hat U(t,0;x,y)$ becomes unitary in the limit $t\to
\infty$.  

Let us now introduce the fluctuating potential $V(r,t)$ and, as we
have already discussed in Sec. 3, assume that
this potential is spatially constant both inside the metallic leads
and inside the dot, and it suffers jumps at the junctions between the
dot and the leads. In other words, we consider three different fluctuating
in time potentials $V_{j}(t)$, where the index $j$ labels
the left lead, the dot and the right lead.

The wave functions are then modified in two ways.
First, they acquire additional phase factors
$\exp\left\{i\int_0^t dt'\, eV_{j}(t')\right\}=\exp\{\varphi_j(t)\}.$
These phase factors can be eliminated by
a gauge transformation, and we will omit them in what follows. In
addition to that, the phase of the wave function changes every time
the electron crosses one of the junctions. The corresponding
additional phase factors acquired by the electron are:
\begin{eqnarray}
{\rm e}^{-i\varphi^+_L} \;\; {\rm left}\;\;{\rm junction,}\;\;{\rm from}\;\;{\rm left}\;\;{\rm to}\;\;{\rm right},
\nonumber\\ 
{\rm e}^{i\varphi^+_L} \;\;{\rm left}\;\;{\rm junction,}\;\;{\rm from}\;\;{\rm right}\;\;{\rm to}\;\;{\rm left},
\nonumber\\
{\rm e}^{-i\varphi^+_R} \;\; {\rm right}\;\;{\rm junction,}\;\;{\rm from}\;\;{\rm left}\;\;{\rm to}\;\;{\rm right},
\nonumber\\
{\rm e}^{i\varphi^+_R} \;\; {\rm right}\;\;{\rm junction,}\;\;{\rm from}\;\;{\rm right}\;\;{\rm to}\;\;{\rm left}.
\label{phasefactors}
\end{eqnarray}
Below we will restrict ourselves to the case $t>0.$
This is sufficient, since the evolution operator at
negative times is just a conjugate operator.
Taking the phase factors (\ref{phasefactors}) into account 
we find $\hat U$ in the presence of fluctuating voltages $V^+_j(t):$
\begin{eqnarray}
\hat U^{\varphi^+}(t,0;x,y)=\hat A(t-x)\hat U(t,0;x,y)\hat A^*(-y)
\label{Uphi}
\end{eqnarray}
where the matrix $\hat A$ reads 
\begin{equation}
\hat A(t)=\left(\begin{array}{cc}
{\rm e}^{i\varphi^+_L(t)}\hat 1 & 0\\
0 & {\rm e}^{-i\varphi^+_R(t)}\hat 1
\end{array}\right).
\label{A}
\end{equation}
The relation (\ref{Uphi}) turns out to be remarkably simple 
since every classical path
-- though being scattered arbitrarily many times inside the dot
and from the outer barriers -- can cross these barriers only
twice: when it enters and leaves the dot. The times of these two
events are related in a trivial way to the total time of the
evolution and to the initial and final coordinates: 
For any $x$ and $y$ the path enters the
dot at a time $-y$ and leaves it at a time $t-x.$ 
We also note that this property
holds only for structures with two barriers. In systems with
three and more barriers no such a simple relation between 
$\hat U^{\varphi^+}(t,0;x,y)$ and $\hat U(t,0;x,y)$ exist.

\section{Effective action}

Let us now use the above expressions for the evolution operator
and derive the effective action of a quantum dot. We start from
the term $S_R$ which is defined by the second term in Eq. (\ref{S1}) as
\begin{eqnarray}
iS_R&=&e\int_0^t dt'\int d^3{\bm r} \big(  G_{V^+,11}(t'-0,t',{\bm r},{\bm r})
\nonumber\\ &&
+\, G_{V^+,22}(t'+0,t',{\bm r},{\bm r})) V^-(t',{\bm r}\big).
\label{SR1}
\end{eqnarray}
Here the space integrals are taken over the dot and both leads.
As we have already discussed, we will adopt the model assuming that
the fluctuating voltage fields inside the leads do not depend on coordinates.
Making use of Eqs. (\ref{Gij},\ref{phi}) let us transform
Eq. (\ref{SR1}) to the following form
\begin{eqnarray}
iS_R&=&2i\int_0^t dt'\int d^3{\bm r}d^3{\bm r}' d^3{\bm r}''\,
U^{\varphi^+}(t'0,{\bm r}{\bm r}')
\nonumber\\ &&\times\,
\rho_0({\bm r}',{\bm r}'')
U^{\varphi^+}(0t',{\bm r}''{\bm r})\dot\varphi^-(t',{\bm r}).
\end{eqnarray}
Let us split this expression by distinguishing each of the three
coordinates ${\bm r}$, ${\bm r}'$ and ${\bm r}''$ to be in the
left lead, in the dot or in the
right lead. Altogether we obtain $3^3=27$ terms. Without
loss of generality this number can be reduced substantially by
means of the following steps. First, we choose the initial density
matrix $\rho_0$ in the form corresponding to zero transmissions of
both junctions. For such a choice initially we have three isolated
systems, in which case the coordinates ${\bm r}'$ and ${\bm r}''$
always belong to the same electrode. This step already reduces the
total number of terms down to nine. Second, we restrict ourselves
to the limit of sufficiently long times, which is of a primary
interest in our problem. In this limit the initial form of the
density matrix inside the dot does not matter at all, electron
transfer and relaxation processes will eventually lead to some
final density matrix in the dot which will be {\it independent} of
the initial one.  Hence, in the interesting limit of long
times our effective action should not depend on the initial
density matrix inside the dot and we can safely exclude the dot
region from the integration over ${\bm r}'$ and ${\bm
  r}''$ extending this integration only at the two reservoirs.
After that we are left with only six terms \cite{FN5}, and the action
$S_R$ takes the form
\begin{eqnarray}
iS_R=2i\sum_{s=l,d,r}\sum_{s'=l,r}\int_0^t dt'\int_{s'} d^3{\bm r}' d^3{\bm r}''\,
\rho_0^{s'}({\bm r}',{\bm r}'')
\nonumber\\ \times
\int_s d^3{\bm r}
U^{\varphi^+}_{ss'}(t'0,{\bm r}{\bm r}')
U^{\varphi^+}_{s's}(0t',{\bm r}''{\bm r})\dot\varphi^-_s(t').
\label{SR2}
\end{eqnarray}

In order to proceed further, we will make use of the unitarity of the
evolution operator. It yields
\begin{eqnarray}
\sum_{s=l,d,r}\int_s d^3{\bm r}
U^{\varphi^+}_{ss'}(t'0,{\bm r}{\bm r}')
U^{\varphi^+}_{s's}(0t',{\bm r}''{\bm r})
\equiv \delta({\bm r}'-{\bm r}'').
\label{unitarity}
\end{eqnarray}
With the aid of this
identity one can express the integrals over ${\bm r}$ in the dot
region through the integrals over the coordinate in the leads,
where an explicit form of the evolution operator (\ref{Uphi}) 
is already available. We find
\begin{eqnarray}
iS_R&=& 2i\int_0^t dt'\,\dot\varphi^-_d(t')\sum_{s'=l,r}\int_{s'} d^3{\bm r}' \,
\rho_0^{s'}({\bm r}',{\bm r}') 
\nonumber\\ &&
+\,2i\sum_{s,s'=l,r}\int_0^t dt'\int_{s'} d^3{\bm r}' d^3{\bm r}''\,
\rho_0^{s'}({\bm r}',{\bm r}'')
\nonumber\\ &&\times\,
\bigg[
\int_l d^3{\bm r}
U^{\varphi^+}_{s'l}(0t',{\bm r}''{\bm r})U^{\varphi^+}_{ls'}(t'0,{\bm r}{\bm r}')
\dot\varphi^-_L(t')
\nonumber\\ &&
-\int_r d^3{\bm r}
U^{\varphi^+}_{s'r}(0t',{\bm r}''{\bm r})U^{\varphi^+}_{rs'}(t'0,{\bm r}{\bm r}')
\dot\varphi^-_R(t')
\bigg],\nonumber
\end{eqnarray}
where $\varphi^-_L=\varphi^-_{l}-\varphi^-_d,$
$\varphi^-_R=\varphi^-_d-\varphi^-_r.$
The first term in this equation is proportional to the total number of
electrons in the leads. Because of the charge neutrality this term should be
exactly canceled by the analogous term from the ion
background. For this reason we will omit this term in our further consideration.

In order to evaluate the remaining terms we switch to the channel
representation. Since we assume that the density matrix in the leads
corresponds to the equilibrium Fermi distribution of electrons, both
this density matrix and the matrix of the phases
$\varphi^-$ are diagonal in the channel space
\begin{equation}
\hat\rho_0(x-y)=\rho_0(x-y)\hat 1, \;
\hat{\dot\varphi}(t')=\left(\begin{array}{cc}
-\hat 1 \dot\varphi_L(t') & 0 \\
 0& \hat 1\dot\varphi_R(t')
\end{array}\right).
\label{rho0}
\end{equation}
The action $S_R$ acquires the following form:
\begin{eqnarray}
iS_R&=&-2i\int dx dy\, {\rm tr}
\bigg[\hat\rho_0(x-y) \int_0^t dt'\int dz
\hat A(-y)
\nonumber\\&& \times\,
\hat U(0,t';y,z)\hat A^*(t-z)
\hat{\dot\varphi}^-(t')\hat A(t-z)
\nonumber\\&& \times\,
\hat U(t',0;z,x)\hat A^*(-x)\bigg].
\label{SR4}
\end{eqnarray}
We also notice that the matrices $\hat{\dot\varphi}(t')$ (\ref{rho0})
and $\hat A$ (\ref{A}) commute, hence $\hat A^*(t-z)$ and $\hat A(t-z)$
drop out. Performing the change of the integration variables
$x\to -x,$ $y\to -y$, we obtain
\begin{equation}
iS_R=-2i\int dx dy\, {\rm tr}
\left[\hat A^*(x)\hat\rho_0(y-x)\hat A(y)\hat B(t,y,x) \right].
\label{SR5}
\end{equation}
Here we defined the matrix
\begin{equation}
\hat B(t,y,x)=\int_0^t dt'\int dz\,
\hat U(0,t';-y,z)\hat{\dot\varphi}^-(t')\hat U(t',0;z,-x).
\end{equation}
With the aid of Eq. (\ref{U0}) this matrix can be evaluated
explicitly. For our purposes it is sufficient to set
$\hat\varphi^-(t)=\hat\varphi^-(0)=0.$ Then we get
\begin{eqnarray}
\hat B(t,y,x)&=& \theta(x)\theta(y)\int_{0}^{t}
dz\,\big[\delta(z-y)\hat\varphi^-(z)\delta(z-x) 
\nonumber\\ &&
-\,\hat S^+(z-y)\hat\varphi^-(z) \hat S(z-x)\big]. \label{B}
\end{eqnarray}
Making use of this expression together with (\ref{A}) and
(\ref{rho0}) one can now directly multiply matrices in Eq.
(\ref{SR5}) and arrive at the final result for $S_R$ presented in
Eq. (\ref{SRcompact}).

Let us now turn to the expression for $S_I$ which is defined by the
last term in Eq. (\ref{S1})
\begin{eqnarray}
S_I&=&e^2 \int_0^t dt' \int_0^t dt''\int d^3{\bm r}' d^3{\bm r}''\,
G_{V^+,12}(t',t'',{\bm r'},{\bm r''})) 
\nonumber\\&& \times\,
 {V}^-(t'',{\bm r}'') G_{V^+,21}(t'',t',{\bm r}'',{\bm r}')) {V}^-(t',{\bm r}').
\nonumber
\end{eqnarray}
The whole consideration is completely analogous to that carried out
above for the term $S_R$. Making use of Eq. (\ref{Gij}), we find
\begin{eqnarray}
S_I=e^2\int_0^t dt_1\int_0^t dt_2\int d^3{\bm x}_1 d^3{\bm x}_2 d^3{\bm y}_1 d^3{\bm y}_2 d^3{\bm z}_1 d^3{\bm z}_2
\nonumber\\ \times\,
U_{\varphi^+}(t_10,{\bm x}_1{\bm y}_1)\rho_0({\bm y}_1,{\bm y}_2)U_{\varphi^+}(0t_2,{\bm y}_2{\bm x}_2)V^-(t_2,{\bm x}_2)
\nonumber\\ \times\,
U_{\varphi^+}(t_20,{\bm x}_2{\bm z}_2)h_0({\bm z}_2,{\bm z}_1)U_{\varphi^+}(0t_1,{\bm z}_1,{\bm x}_1)V^-(t_1,{\bm x}_1).
\nonumber
\end{eqnarray}
where $h_0({\bm z}_2,{\bm z}_1)=1-\rho_0({\bm z}_2,{\bm z}_1)$. As
before we identify the contributions containing the initial
density matrix in the dot. Such terms can again be omitted in the
interesting limit of long times. We again apply the unitarity
condition (\ref{unitarity}) and explicitly introduce the channel
indices. Then after some straightforward manipulations we obtain
\begin{eqnarray}
S_I=\int_0^t dt_1 \int_0^t dt_2\int dx_1dx_2  dy_1dy_2 dz_1 dz_2 
\nonumber\\ \times\,
{\rm tr}\bigg\{
\hat U(0,t_1;z_1,x_1)\hat{\dot\varphi}^-(t_1)\hat U(t_1,0;x_1,y_1)\hat A^*(-y_1)
\nonumber\\ 
\times \hat \rho_0(y_1-y_2)\hat A(-y_2)
\hat U(0,t_2;y_2,x_2)\hat{\dot\varphi}^-(t_2)
\nonumber\\
\times
\hat U(t_2,0;x_2,z_2)\hat A^*(-z_2)\hat h_0(z_2-z_1)\hat A(-z_1)
\bigg\}.\nonumber
\end{eqnarray}
With the aid of the matrix (\ref{B}) this expression can be
transformed further and eventually takes the form
\begin{eqnarray}
S_I&=& \int dy_1dy_2\int dz_1 dz_2 {\rm tr}\bigg\{
\hat B(z_1,y_1)
 \hat A^*(y_1)\hat \rho_0(y_2-y_1)
\nonumber\\ &&\times\,
\hat A(y_2)
\hat B(y_2,z_2) \hat A^*(z_2)\hat h_0(z_1-z_2)\hat A(z_1)
\bigg\}. \label{ddddd}
\end{eqnarray}
Multiplying matrices in Eqs. (\ref{ddddd}) we arrive at the final
expressions for $S_I$ defined in Eq. (\ref{SIcompact}).

\section{Averaging over circular ensemble of $S-$matrices}

Here we will evaluate the following average 
\begin{equation}
W={\rm tr}\langle\hat A\hat S^\dagger(E_1)\hat B\hat S(E_2)\hat C\hat S^\dagger(E_3)\hat D\hat S(E_4)\rangle.
\label{W1}
\end{equation}
Let us use the representation (\ref{ST}) and split the $S-$matrix
into two parts 
\begin{equation}
\hat S(E)=\hat R+\delta\hat S(E), \;\;
\delta\hat S(E)=\hat T'[1-\hat U(E)]^{-1}\hat U(E)\hat T.
\label{split}
\end{equation}
One can show that $\langle\delta\hat S(E)\rangle=0.$
At any given energy $E$ the matrix $\hat U(E)$
belongs to the circular ensemble. 

In a semi-classical limit averaging of the products
of such matrices taken at the same energy can be
performed with the aid of the diagram technique \cite{BB}.
Here we need a more general version of this technique which would
also allow to average the products of matrices taken at 
different energies.
Appropriate modifications of the rules can be formulated with the aid
of the results \cite{Brouwer}. 
In short, using the terminology \cite{BB}, every $U-$cycle involving two matrices
$\hat U(E_1)$ and $\hat U^\dagger(E_2)$ carries an additional factor
$M/(M-iE_{12}t_0)$ as compared to the situation \cite{BB}. 
Here $M=N_L+N_R$, where $N_{L(R)}$ is the number of channels in the
left (right) barrier/lead, $t_0=2\pi /\delta$ and $E_{ij}=E_i-E_j.$ 
An $U-$cycle involving four matrices $\hat U^\dagger(E_1),$
$\hat U(E_2),$ $\hat U^\dagger(E_3)$ and $\hat U(E_4)$
gives an additional factor
$$
\frac{M^3(M-i(E_{23}+E_{41})t_0)}
{(M-iE_{41}t_0)(M-iE_{21}t_0)(M-iE_{23}t_0)(M-iE_{43}t_0)}.
$$

Higher order $U-$cycles do not contribute in the semi-classical limit
$M\to \infty,$ and we will not consider them here.

Employing Eq. (\ref{split}) one can express the average (\ref{W1})
as a sum of 10 non-vanishing terms.
The first of them is trivial and equals
${\rm tr}(\hat A\hat R^\dagger\hat B\hat R\hat C\hat R^\dagger\hat D\hat R).$
Several terms are bilinear in the matrices $\delta\hat S$
and $\delta\hat S^\dagger.$ Such terms are evaluated with the aid of the
formula
\begin{eqnarray}
{\rm tr}\langle \hat A\delta\hat S(E_1)\hat B\delta\hat S^\dagger(E_2)\rangle =
\frac{({\rm tr}\hat {T'}^\dagger\hat A\hat T')({\rm tr}\hat T\hat B\hat T^\dagger)}
{M-{\rm tr}\hat R'\hat{R'}^\dagger -iE_{12}t_0}
\nonumber\\
= \frac{2({\rm tr}\hat {T'}^\dagger\hat A\hat T')({\rm tr}\hat T\hat B\hat T^\dagger)}
{g(1-iE_{12}\tau_D)}.
\label{two}
\end{eqnarray}
Here we have used the identity $M-{\rm tr}\hat R'\hat{R'}^\dagger={\rm tr}\hat T\hat{T}^\dagger=g,$
and defined the dwell time $\tau_D=2t_0/g.$
Eq. (\ref{two}) is a direct generalization of the analogous
formula \cite{BB} derived for the case $E_1=E_2$. This formula allows
to evaluate the averages $u_\omega^{ij}$ (\ref{uuu}). For example,
$u^{RL}_\omega={\rm tr}
\langle\hat t^\dagger(E)\hat t(E+\omega)\rangle=
{\rm tr}\langle\hat C_1\hat S^\dagger(E)\hat C_2\hat S(E+\omega)\rangle,$
where we defined the matrices
\begin{equation}
\hat C_1=\left(\begin{array}{cc}\hat 1 & 0 \\ 0 & 0 \\\end{array}\right), \;\;
\hat C_2=\left(\begin{array}{cc} 0 & 0 \\ 0 & \hat 1 \\\end{array}\right).
\label{Cj}
\end{equation}
Employing Eq. (\ref{two}) we arrive at (\ref{uav}).

The terms involving three matrices $\delta\hat S$ and/or $\delta\hat S^\dagger$
are slightly more complicated. They read
\begin{widetext}
\begin{eqnarray}
{\rm tr}\langle\hat A\delta\hat S(E_1)\hat B\delta\hat S(E_2)\hat C\delta\hat S^\dagger(E_3)\rangle
&=&\frac{4({\rm tr}\hat {T'}^\dagger\hat A\hat T')({\rm tr}\hat T\hat C\hat T^\dagger)
({\rm tr}\hat T\hat B\hat T'\hat{R'}^\dagger)}
{g^2(1-iE_{13}\tau_D)
(1-iE_{23}\tau_D)},
\nonumber\\
{\rm tr}\langle\hat A\delta\hat S(E_1)\hat B\delta\hat S^\dagger(E_2)
\hat C\delta\hat S^\dagger(E_3)\rangle
&=&\frac{4({\rm tr}\hat {T'}^\dagger\hat A\hat {T'})({\rm tr}\hat T\hat B\hat T^\dagger)
({\rm tr}\hat R'\hat {T'}^\dagger\hat C\hat{T}^\dagger)}
{g^2(1-iE_{12}\tau_D)
(1-iE_{13}\tau_D)}.
\label{three}
\end{eqnarray}
The derivation of Eqs. (\ref{three}) is straightforward, and we will
not discuss the corresponding details here. 

\begin{figure}
\begin{tabular}{ccc}
\includegraphics[width=4.5cm]{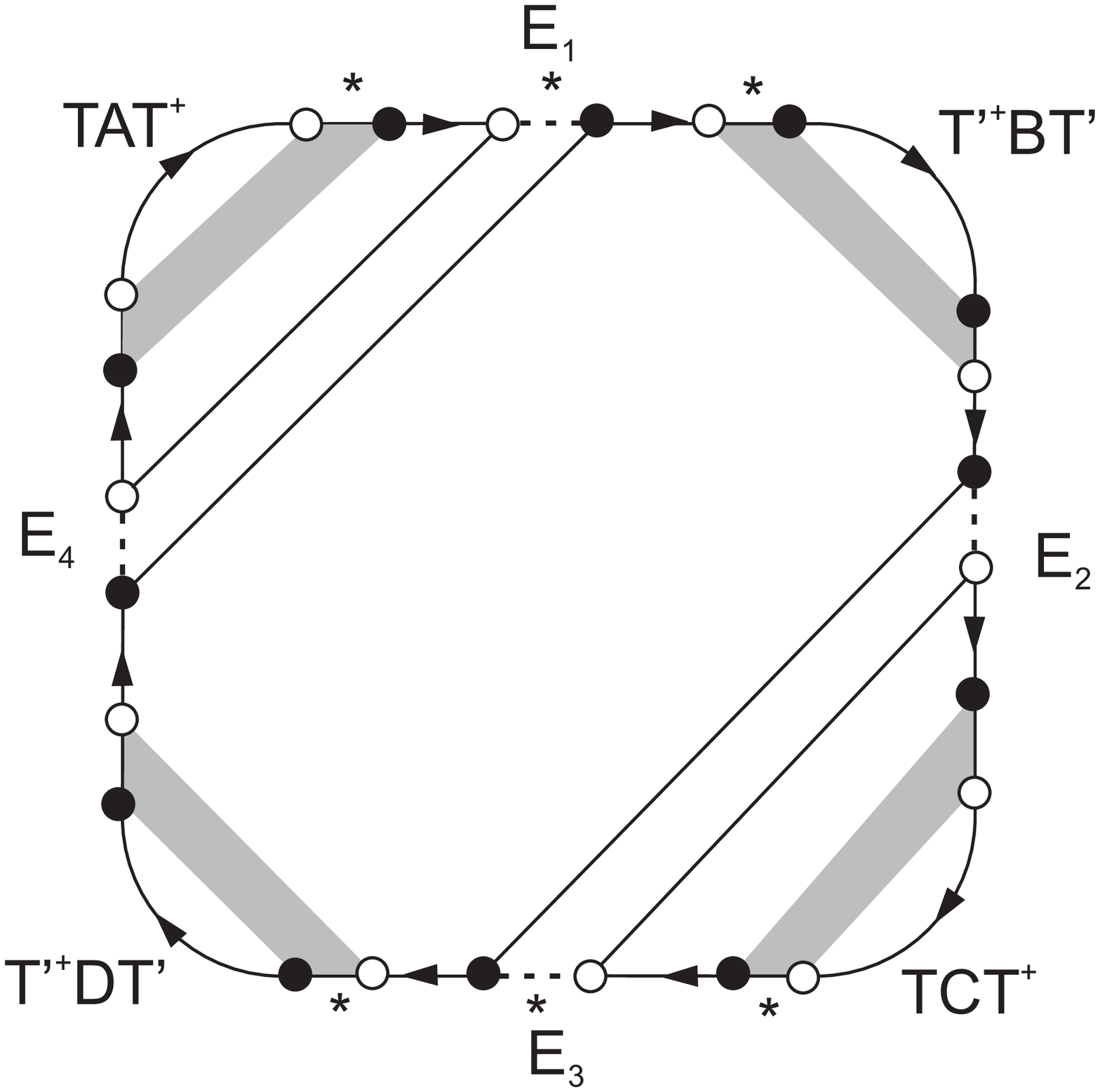} &
\includegraphics[width=4.5cm]{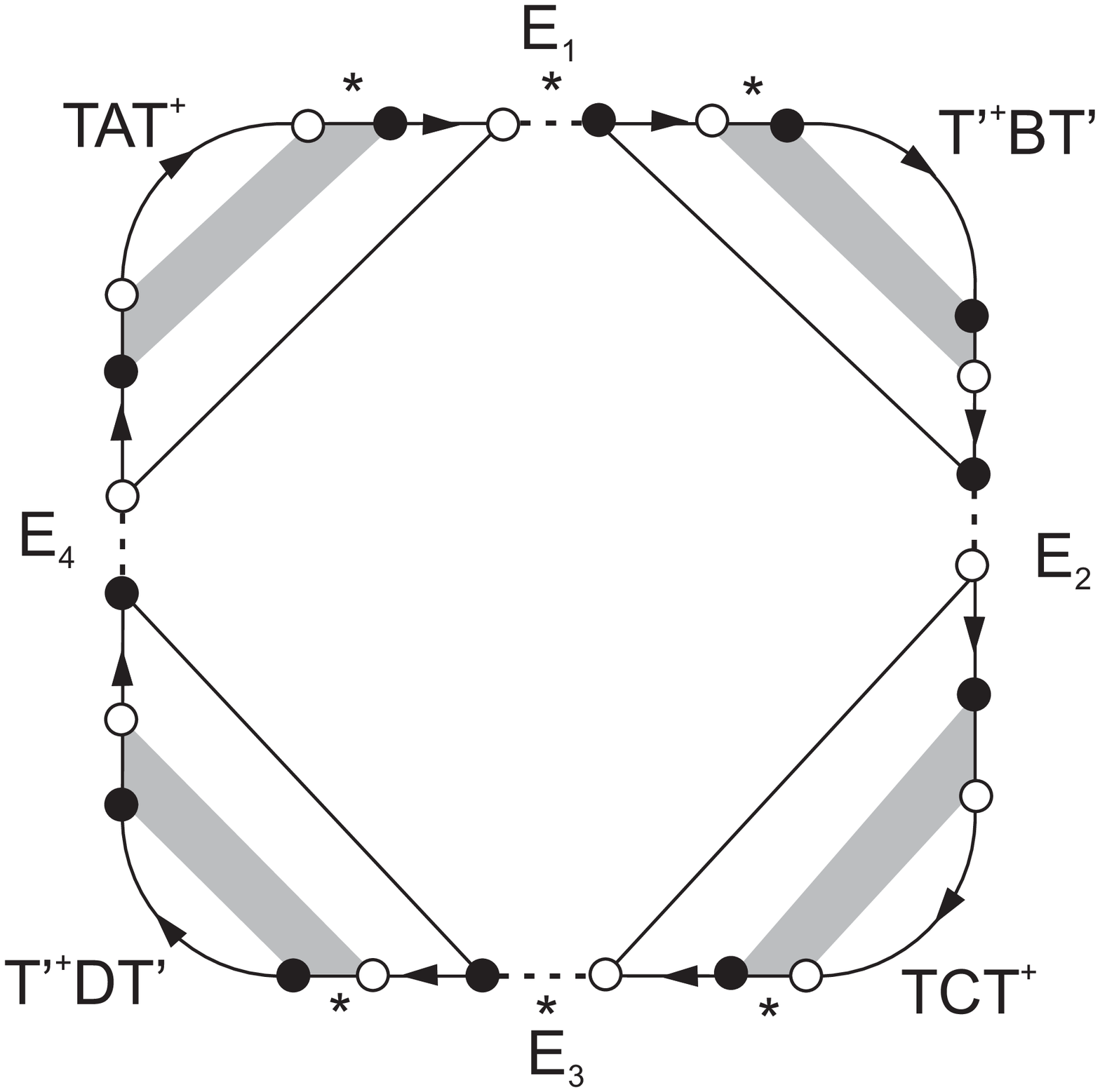} &
\includegraphics[width=4.5cm]{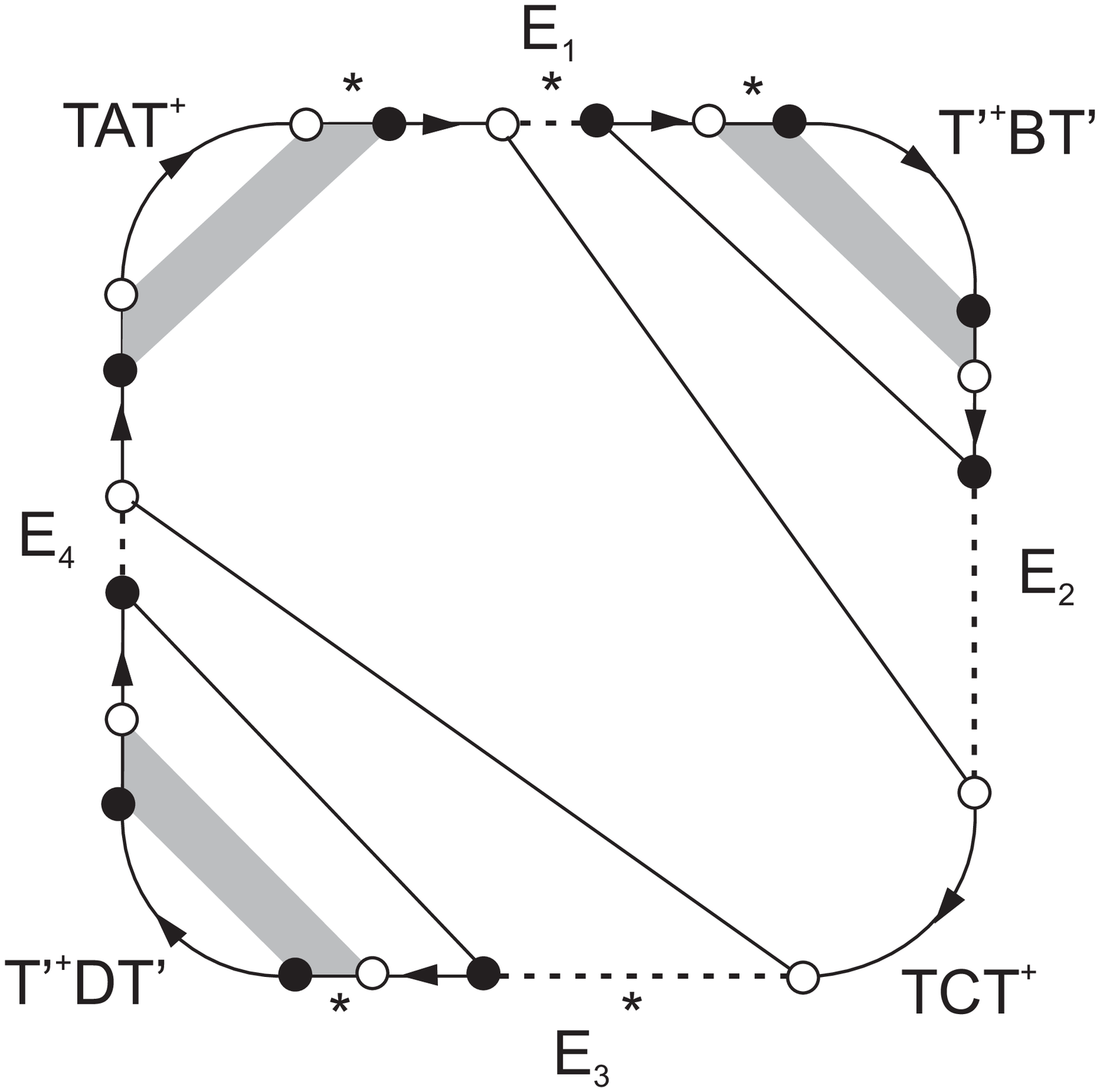} \\ 
$D_1$ & $D_2$  & $D_3$ \\
\includegraphics[width=4.5cm]{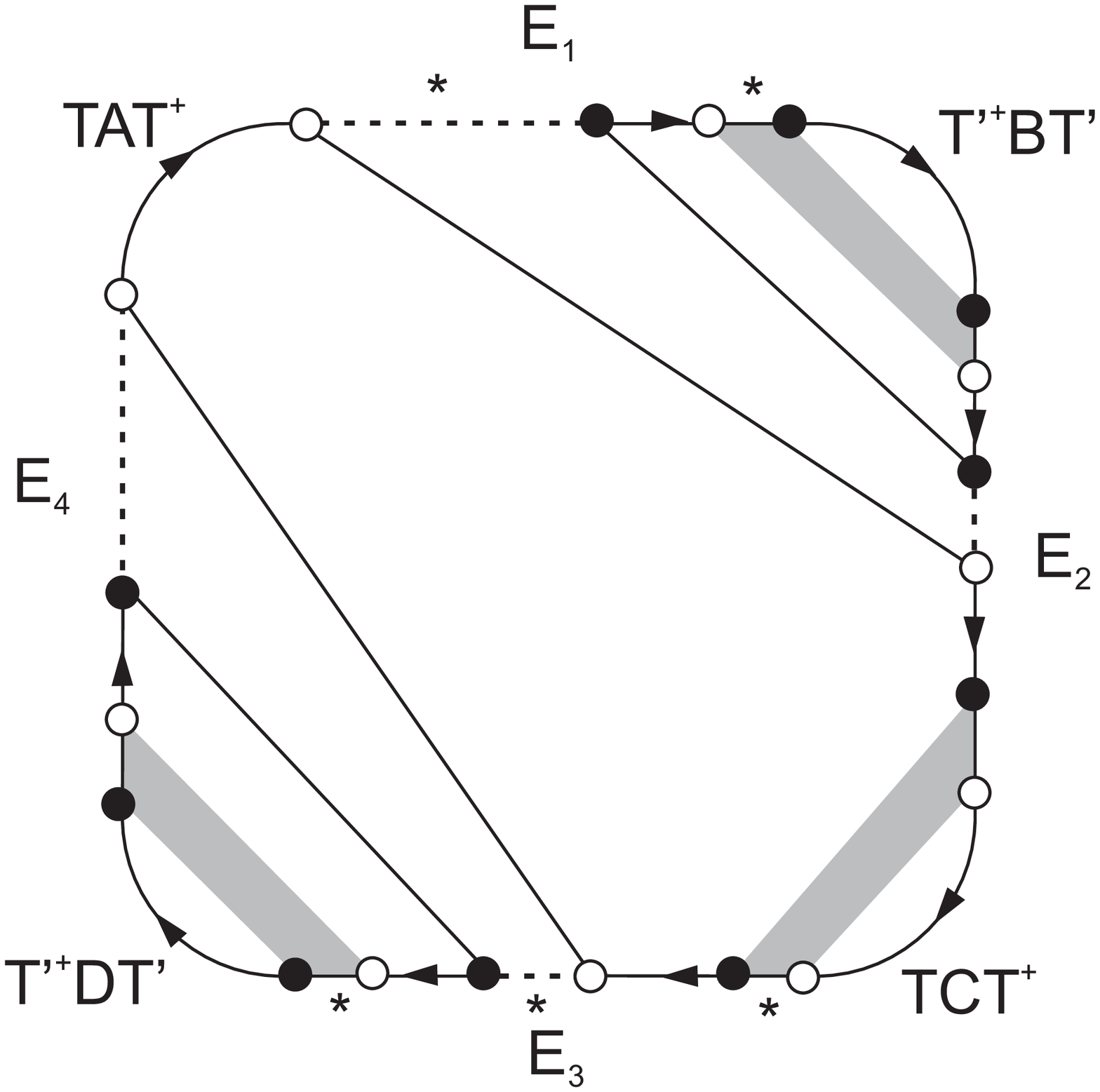} &
\includegraphics[width=4.5cm]{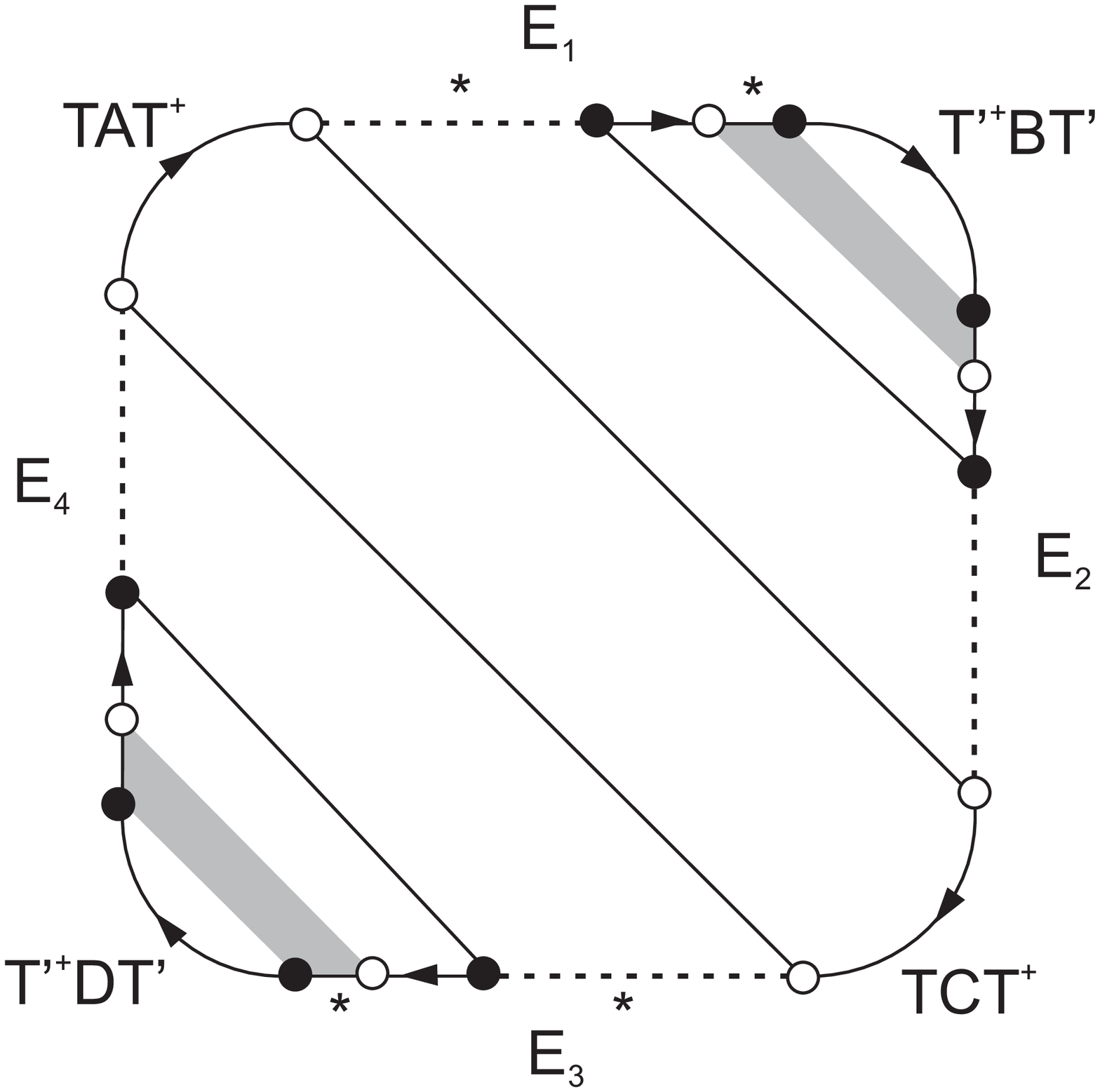} \\
$D_4$ & $D_5$ \\
\end{tabular}
\caption{Five diagrams contributing to 
${\rm tr}\langle\hat A\delta\hat S^\dagger(E_1)\hat B\delta\hat S(E_2)\hat C\delta\hat S^\dagger(E_3)\hat D\delta\hat S(E_4)\rangle$.
We refer to  Ref. \onlinecite{BB} for the definition of the graphical
objects presented here.}
\label{Diag}
\end{figure}
What remains is to evaluate the average
\begin{equation}
w={\rm tr}\langle\hat A\delta\hat S^\dagger(E_1)\hat B\delta\hat S(E_2)\hat C\delta\hat S^\dagger(E_3)\hat D\delta\hat S(E_4)\rangle.
\label{four1}
\end{equation}
This average is given by the combination of five diagrams 
shown in Fig. \ref{Diag}:
\begin{equation}
w= D_1+D_2+D_3+D_4-D_5.
\end{equation}
The diagram $D_5$ is subtracted in order to compensate for double counting
of certain graphs in the diagrams $D_3$ and $D_4.$
The contributions of the diagrams read
\begin{eqnarray}
D_1&=&\frac{({\rm tr}\hat F_A)({\rm tr}\hat F_C)({\rm tr}\hat F_B\hat F_D)}{(M-iE_{41}t_0)(M-iE_{23}t_0)}, \;\;
D_2=-\frac{({\rm tr}\hat F_A)({\rm tr}\hat F_B)({\rm tr}\hat F_C)({\rm tr}\hat F_D)(M-i(E_{23}+E_{41})t_0)}
{(M-iE_{41}t_0)(M-iE_{21}t_0)(M-iE_{23}t_0)(M-iE_{43}t_0)},
\nonumber\\
D_3&=&\frac{({\rm tr}\hat F_A\hat T\hat C\hat T^\dagger)({\rm tr}\hat F_B)({\rm tr}\hat F_D)}
{(M-iE_{21}t_0)(M-iE_{43}t_0)},\;
D_4=\frac{({\rm tr}\hat T\hat A\hat T^\dagger\hat F_C)({\rm tr}\hat F_B)({\rm tr}\hat F_D)}
{(M-iE_{21}t_0)(M-iE_{43}t_0)},
\;
D_5=\frac{({\rm tr}\hat T\hat A\hat T^\dagger\hat T\hat C\hat T^\dagger )({\rm tr}\hat F_B)({\rm tr}\hat F_D)}
{(M-iE_{21}t_0)(M-iE_{43}t_0)}. 
\label{Di}
\end{eqnarray}
Here we have introduced the ladder blocks defined in Fig. \ref{Ladder}:
\begin{eqnarray}
\hat F_A&=&\hat T\hat A\hat T^\dagger +\sum_{n=0}^\infty ({\rm tr}\hat T\hat A\hat T^\dagger)\hat R'\hat {R'}^\dagger
\frac{({\rm tr}\hat R'\hat {R'}^\dagger)^n}{(M-iE_{41}t_0)^{n+1}} 
=\hat T\hat A\hat T^\dagger + \frac{2{\rm tr}\hat T\hat A\hat T^\dagger}{g(1-iE_{41}\tau_D)}\hat R'\hat {R'}^\dagger,
\nonumber\\
\hat F_B&=&\hat {T'}^\dagger\hat B\hat T' + \frac{2{\rm tr}\hat {T'}^\dagger\hat B\hat T'}{g(1-iE_{21}\tau_D)}\hat R'\hat {R'}^\dagger,\;\;
\hat F_C=\;\hat T\hat C\hat T^\dagger + \frac{2{\rm tr}\hat T\hat C\hat T^\dagger}{g(1-iE_{23}\tau_D)}\hat R'\hat {R'}^\dagger,
\hat F_D=\hat {T'}^\dagger\hat D\hat T' + \frac{2{\rm tr}\hat {T'}^\dagger\hat D\hat T'}{g(1-iE_{43}\tau_D)}\hat R'\hat {R'}^\dagger.
\nonumber
\end{eqnarray}

Eventually we arrive at an important -- though rather lengthy -- formula
\begin{eqnarray}
W&=&{\rm tr}\hat A\hat R^\dagger\hat B\hat R\hat C\hat R^\dagger\hat D\hat R
+\frac{2({\rm tr}\hat {T'}^\dagger\hat B\hat T')
({\rm tr}\hat T\hat C\hat R^\dagger\hat D\hat R\hat A\hat T^\dagger)}
{g (1-iE_{21}\tau_D)}
+\frac{2({\rm tr}\hat {T'}^\dagger\hat B\hat R\hat C\hat R^\dagger\hat D\hat T')
({\rm tr}\hat T\hat A\hat T^\dagger)}
{g (1-iE_{41}\tau_D)}
\nonumber\\ &&
+\,\frac{2({\rm tr}\hat {T'}^\dagger\hat D\hat R\hat A\hat R^\dagger\hat B\hat T')
({\rm tr}\hat T\hat C\hat T^\dagger)}
{g (1-iE_{23}\tau_D)}
+\frac{2({\rm tr}\hat {T'}^\dagger\hat D\hat T')
({\rm tr}\hat T\hat A\hat R^\dagger\hat B\hat R\hat C\hat T^\dagger)}
{g (1-iE_{43}\tau_D)}
\nonumber\\ &&
+\,\frac{4({\rm tr}\hat{T'}^\dagger\hat B\hat T')({\rm tr}\hat T\hat C\hat T^\dagger)
({\rm tr}\hat R'\hat{T'}^\dagger\hat D\hat R\hat A\hat T^\dagger)}
{g^2(1-iE_{21}\tau_D)(1-iE_{23}\tau_D)}
+\frac{4({\rm tr}\hat{T'}^\dagger\hat B\hat T')({\rm tr}\hat T\hat A\hat T^\dagger)
({\rm tr}\hat T\hat C\hat R^\dagger\hat D\hat T'\hat{R'}^\dagger)}
{g^2(1-iE_{21}\tau_D)(1-iE_{41}\tau_D)}
\nonumber\\ &&
+\,\frac{4({\rm tr}\hat{T'}^\dagger\hat D\hat T')({\rm tr}\hat T\hat A\hat T^\dagger)
({\rm tr}\hat R'\hat {T'}^\dagger\hat B\hat R\hat C\hat T^\dagger)}
{g^2(1-iE_{41}\tau_D)(1-iE_{43}\tau_D)}
+\frac{4({\rm tr}\hat{T'}^\dagger\hat D\hat T')({\rm tr}\hat T\hat C\hat T^\dagger)
({\rm tr}\hat T\hat A\hat R^\dagger\hat B\hat T'\hat{R'}^\dagger)}
{g^2(1-iE_{43}\tau_D)(1-iE_{23}\tau_D)}
\nonumber\\&&
+\, \frac{4({\rm tr}\hat T\hat A\hat T^\dagger)({\rm tr}\hat T\hat C\hat T^\dagger){\rm tr}(\hat {T'}^\dagger\hat B\hat T'\hat {T'}^\dagger\hat D\hat T')}
{g^2(1-iE_{41}\tau_D)(1-iE_{23}\tau_D)}+
\frac{4({\rm tr}\hat{T'}^\dagger\hat B\hat T')({\rm tr}\hat{T'}^\dagger\hat D\hat T')({\rm tr}\hat T\hat A\hat T^\dagger\hat T\hat C\hat T^\dagger)}
{g^2(1-iE_{21}\tau_D)(1-iE_{43}\tau_D)}
\nonumber\\ &&
+\,\frac{8({\rm tr}\hat T\hat A\hat T^\dagger)({\rm tr}\hat T\hat C\hat T^\dagger)({\rm tr}\hat {T'}^\dagger\hat D\hat T')({\rm tr}\hat {T'}^\dagger\hat B\hat T'\hat {R'}^\dagger\hat R')}
{g^3(1-iE_{43}\tau_D)(1-iE_{41}\tau_D)(1-iE_{23}\tau_D)}
+\frac{8({\rm tr}\hat T\hat A\hat T^\dagger)({\rm tr}\hat T\hat C\hat T^\dagger)({\rm tr}\hat {T'}^\dagger\hat B\hat T')({\rm tr}\hat {T'}^\dagger\hat D\hat T'\hat {R'}^\dagger\hat R')}
{g^3(1-iE_{21}\tau_D)(1-iE_{41}\tau_D)(1-iE_{23}\tau_D)}
\nonumber\\&&
+\,\frac{8({\rm tr}\hat{T'}^\dagger\hat B\hat T')({\rm tr}\hat{T'}^\dagger\hat D\hat T')({\rm tr}\hat T\hat A\hat T^\dagger)({\rm tr}\hat T\hat C\hat T^\dagger\hat R'\hat{R'}^\dagger)}
{g^3(1-iE_{21}\tau_D)(1-iE_{43}\tau_D)(1-iE_{41}\tau_D)}
+\frac{8({\rm tr}\hat{T'}^\dagger\hat B\hat T')({\rm tr}\hat{T'}^\dagger\hat D\hat T')({\rm tr}\hat T\hat C\hat T^\dagger)({\rm tr}\hat T\hat A\hat T^\dagger\hat R'\hat{R'}^\dagger)}
{g^3(1-iE_{21}\tau_D)(1-iE_{43}\tau_D)(1-iE_{23}\tau_D)}
\nonumber\\ &&
-\,\frac{16({\rm tr}\hat T\hat A\hat T^\dagger)({\rm tr}\hat T\hat C\hat T^\dagger)({\rm tr}\hat {T'}^\dagger\hat B\hat T')({\rm tr}\hat {T'}^\dagger\hat D\hat T')
(M-{\rm tr}\hat R'\hat{R'}^\dagger\hat R'\hat{R'}^\dagger -i(E_{23}+E_{41})t_0)}
{g^4(1-iE_{21}\tau_D)(1-iE_{43}\tau_D)(1-iE_{41}\tau_D)(1-iE_{23}\tau_D)}.
\label{W}
\end{eqnarray}

\begin{figure}
\includegraphics[width=16cm]{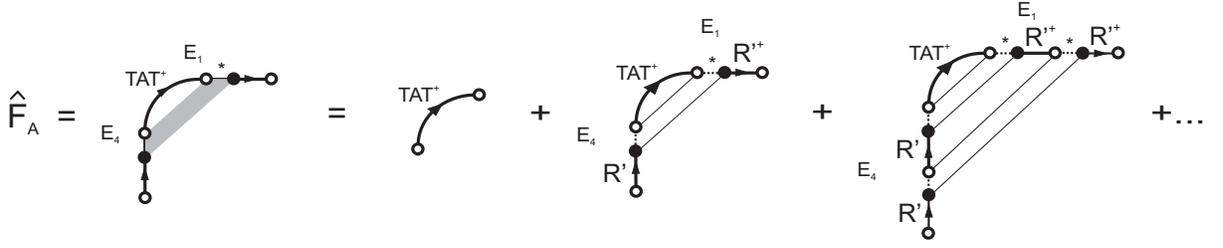} 
\caption{The ladder block $\hat F_A$ appearing in the diagrams shown in Fig \ref{Diag}.
The blocks $\hat F_B,$ $\hat F_C$ and $\hat F_D$ are defined analogously.}
\label{Ladder}
\end{figure}

\end{widetext}

One can demonstrate that the result (\ref{W}) is consistent with the 
unitarity of the $S-$matrix. For this purpose let us put
$E_4=E_3$ and $\hat D=1.$ Due to the unitarity of $S-$matrices of the barriers 
almost all of the terms cancel out and the result
reduces to
\begin{eqnarray}
W&=&
{\rm tr}\hat A\hat R^\dagger\hat B\hat R\hat C
+\frac{2({\rm tr}\hat {T'}^\dagger\hat B\hat T')({\rm tr}\hat T\hat C\hat A\hat T^\dagger)}
{g(1-iE_{21}\tau_D)}
\nonumber\\
&=& {\rm tr}\langle\hat A\hat S^\dagger(E_1)\hat B\hat S(E_2)\hat C\rangle.
\end{eqnarray}
The last equation is a consequence of Eq. (\ref{two}). 
This is just the expression expected from the definition (\ref{W1})
and the unitarity of the $S-$matrix. 

With the aid of
Eq. (\ref{W}) one can find the averages $\tilde v^{ij}_\omega$
defined in Eq. (\ref{dvij}). 
We transform them to the form 
\begin{eqnarray}
\tilde v^{LL}_\omega =
{\rm tr}\langle\hat C_1\hat S^\dagger(E)\hat C_1\hat S(E+\omega)\hat C_2\hat S^\dagger(E+\omega)\hat C_1\hat S(E)\rangle
\hspace{0.5cm}
\nonumber\\ 
\tilde v^{RL}_\omega = {\rm tr}\langle\hat C_1\hat S^\dagger(E)\hat C_2\hat S(E+\omega)\hat C_1\hat S^\dagger(E+\omega)\hat C_1\hat S(E)\rangle
\hspace{0.5cm}
\label{vijav0}
\end{eqnarray}
and similarly  for $\tilde v^{RR}_\omega$ and 
$\tilde v^{LR}_\omega$. Making use of Eq. (\ref{W}) we arrive at Eqs. (\ref{vijav}).
Analogously, we derive the expressions (\ref{Dij11}).

\section{General expressions for the current}
In Sec. 4A we focused on the expression for the current in the
limit of zero external impedance $Z_S \to 0$. Our analysis also allows to
establish more general expressions valid for arbitrary $Z_S(\omega )$.
We obtain
\begin{equation}
I(V)=I_0(V)+\delta I_{LR}+\delta I_{LL}+\delta I_{RR},
\end{equation}
where
\begin{widetext}
\begin{eqnarray}
I_0(V)&=&2e\int dx\,\rho_0(x)\,{\cal T}(x)\,\left({\rm e}^{-F_{LL}(x)}\,{\rm e}^{ieV_Lx}-{\rm e}^{-F_{RR}(x)}\,{\rm e}^{-ieV_Rx}\right),
\label{I03}
\end{eqnarray}
\begin{eqnarray}
\delta I_{LR}&=&2e\,{\rm Im}\int dx\int  dy_1dy_2 dz_1 dz_2\;
[\rho_0(y_2-y_1)h_0(z_1-z_2)+\rho_0(y_1-y_2)h_0(z_2-z_1)]
\nonumber\\ &&\times\,
{\cal F}_{LR}(y_1,y_2,z_1,z_2){\rm e}^{-i[eV_R(y_1-y_2)+eV_L(z_1-z_2)]}
{\rm Tr}\left\{
\hat r^\dagger(x+z_1)\hat t'(x+y_1)\right.
\nonumber\\ &&\left.\times\,
\big[(K_{LR}(-y_1)+K_{LL}(-z_1))
\hat {t'}^\dagger(y_2)\hat r(z_2)
-(K_{RR}(-y_1)+K_{RL}(-z_1))\hat {r'}^\dagger(y_2)\hat t(z_2)\big]
\right\},
\label{ILR3}
\end{eqnarray}
\begin{eqnarray}
\delta I_{LL}&=&-2e\,{\rm Im}\int dx\int dy_1dy_2 dz_1 dz_2\;
\rho_0(y_1-y_2)h_0(z_2-z_1)
{\cal F}_{LL}(y_1,y_2,z_1,z_2){\rm e}^{i[eV_L(y_1-y_2-z_1+z_2)]}
\nonumber\\ &&\times\,
{\rm Tr}\big\{
\hat t^\dagger(x+z_1)\hat t(x+y_1)
\big[(K_{LL}(-y_1)-K_{LL}(-z_1))
\big(\delta(y_2)\delta(z_2)\hat 1-\hat r^\dagger(y_2)\hat r(z_2)\big)
\nonumber\\ &&
+\,(K_{RL}(-y_1)-K_{RL}(-z_1))\hat t^\dagger(y_2)\hat t(z_2)\big]
\big\},
\label{ILL3}
\end{eqnarray}
\begin{eqnarray}
\delta I_{RR}&=&2e\, {\rm Im}\int dx\int  dy_1dy_2 dz_1 dz_2\;
\rho_0(y_1-y_2)h_0(z_2-z_1)
{\cal F}_{RR}(y_1,y_2,z_1,z_2){\rm e}^{-i[eV_R(y_1-y_2-z_1+z_2)]}
\nonumber\\ &&\times\,
{\rm Tr}\big\{
\hat {t'}^\dagger(x+z_1)\hat t'(x+y_1)
\big[(K_{RR}(-y_1)-K_{RR}(-z_1))
\big(\delta(y_2)\delta(z_2)\hat 1-\hat {r'}^\dagger(y_2)\hat r'(z_2)\big)
\nonumber\\&&
+\,
(K_{LR}(-y_1)-K_{LR}(-z_1))\hat {t'}^\dagger(y_2)\hat t'(z_2)\big]
\big\}.
\label{IRR3}
\end{eqnarray}
Here we defined
\begin{eqnarray}
{\cal F}_{LR}(y_1,y_2,z_1,z_2)&=&
{\rm e}^{-F_{RR}(y_1-y_2)-F_{LL}(z_1-z_2)+F_{LR}(y_1-z_1)+F_{LR}(y_2-z_2)
-F_{LR}(y_1-z_2)-F_{LR}(y_2-z_1)},
\label{calFLR}\\
{\cal F}_{LL}(y_1,y_2,z_1,z_2)&=&
{\rm e}^{-F_{LL}(y_1-y_2)-F_{LL}(z_1-z_2)-F_{LL}(y_1-z_1)-F_{LL}(y_2-z_2)
+F_{LL}(y_1-z_2)+F_{LL}(y_2-z_1)},
\label{calFLL}
\end{eqnarray}
\begin{eqnarray}
F_{ij}(x)&=&\int\frac{d\omega}{2\pi} \,{\rm Im}K_{ij}(\omega)\coth\frac{\omega}{2T}\,(1-\cos\omega x)
\nonumber\\ &&
+\,\int\frac{d\omega}{(2\pi)^2}\left(\sum_{\pm}(\omega\pm
eV)\coth\frac{\omega\pm eV}{2T}-2\omega\coth\frac{\omega}{2T}\right)(1-\cos\omega x)
\sum_{k,n=L,R} K^*_{ik}(\omega)\delta v_{kn}(\omega)K_{nj}(\omega).
\label{FS}
\end{eqnarray}
$\delta v_{kn}(\omega)$ are defined in Eq. (\ref{dvij}).
The function ${\cal F}_{RR}$ is obtained by substituting $F_{LL}(x) \to 
F_{RR}(x)$ into Eq. (\ref{calFLL}).
\end{widetext}


\begin{thebibliography}{}
\bibitem{FN0} Within this approach an effective energy dependence of 
(renormalized) transmissions $T_n$ can only be generated by 
electron-electron interactions \cite{Naz2,BN}.
\bibitem{FN4} The authors \cite{Basel} have fitted their 
  data to the power law $G \propto T^p$ with $p \sim 0.1\div 0.2$.
In the temperature range of about one decade such a dependence can
hardly be distinguished from a logarithmic one.
\bibitem{FN3} The same result follows if one repeats the calculation
\cite{many3} for an array of $M\times N$ ($N \gg 1$) independent
coherent scatterers with the Fano factor $\tilde \beta$.
\bibitem{FN7} Following the authors \cite{PC2} we assume that
  electrons are mainly conducted by the outer shell of the nanotube.
In this case one can define $M=L_{\rm tr}/L_{\rm s}$ and $L_{\rm tr}=\pi
  d$, where $d$ is the nanotube diameter.
\bibitem{FN5} We have also verified our general arguments by directly
calculating {\it all} the contributions to the effective action. 
This calculation also demonstrates that in the long time limit
our results are insensitive to the terms omitted here.
\end{thebibliography}
\end{document}